\documentclass{article}
\usepackage{multicol}
\usepackage{graphicx} % Required for inserting images
\usepackage{subcaption} % Required for subimages
\usepackage{amsmath} % Math
\usepackage{textgreek} % Greek
\usepackage{xcolor}
\usepackage{gensymb}
\usepackage{bm}
\usepackage{geometry} % Adjust margins
\usepackage{float}
\usepackage{placeins}
\usepackage{comment}
\usepackage{setspace} % Adjust text spacing
\usepackage{enumitem} % Make lists
\usepackage{authblk} % Author affiliations
\usepackage[bottom]{footmisc} % Make footnotes at bottom of page
\usepackage{caption}
\usepackage[
backend=biber,
style=chem-acs,
sorting=none,
safeinputenc
]{biblatex}
\usepackage{titlesec}
\titlespacing*{\section} {0pt}{2ex}{0ex}
\titlespacing*{\subsection} {0pt}{2ex}{0ex}

\titlespacing*{\subsubsection} {0pt}{2ex}{0ex}
\begin{document}

\renewcommand\Affilfont{\itshape\small}
\renewcommand{\thefootnote}{\fnsymbol{footnote}}

\title{\textbf{Disentangling the ferroelectric phases of epitaxial hafnia}}

\author[1]{Johanna van Gent$^{*}$}
\author[1]{Ewout van der Veer}
\author[1]{Yulei Li}
\author[2]{Daniel A. Chaney\footnote{daniel.chaney@esrf.fr}}
\author[1,3]{Beatriz Noheda\footnote{b.noheda@rug.nl}}

\affil[1]{Zernike Institute for Advanced Materials, University of Groningen, Groningen, The Netherlands}
\affil[2]{The European Synchrotron Radiation Facility (ESRF), Grenoble, France}
\affil[3]{CogniGron centre, University of Groningen, Groningen, The Netherlands}
\affil[*]{These authors contributed equally to this work.}

\date{}
\maketitle

\renewcommand\thesection{\gulliver{\arabic{section}}}
\captionsetup[figure]{labelfont={bf},labelformat={default},labelsep=period,name={Fig.}}
\captionsetup[table]{labelfont={bf},labelformat={default},labelsep=period,name={Table}}

\vspace*{-3mm}
\noindent \textbf{Since its discovery, ferroelectric hafnia has been extensively studied due to its CMOS-compatibility and ability to remain polarized at sub-10 nm thicknesses. The ferroelectric behaviour is generally attributed to a polar orthorhombic "OIII" phase. However, a second polar phase with rhombohedral symmetry (R-phase) has also been reported in epitaxial films. The nature of the R-phase remains disputed due to the subtle differences with the OIII-phase when probed by standard thin film characterisation techniques. Given the functional properties of ferroelectrics are crucially determined by the crystal symmetry, resolving this matter is imperative. In this work, we settle the controversy through extensive 3D reciprocal space surveys made possible via synchrotron-based grazing incidence diffraction from epitaxial films of both phases. These experiments, together with direct comparison of their temperature dependence and electrical responses, conclusively establish them as two distinct phases and provide insight into their key characteristics.}

\begin{multicols}{2}

\begin{figure*}[!ht]
    \centering
    \includegraphics[width=\linewidth]{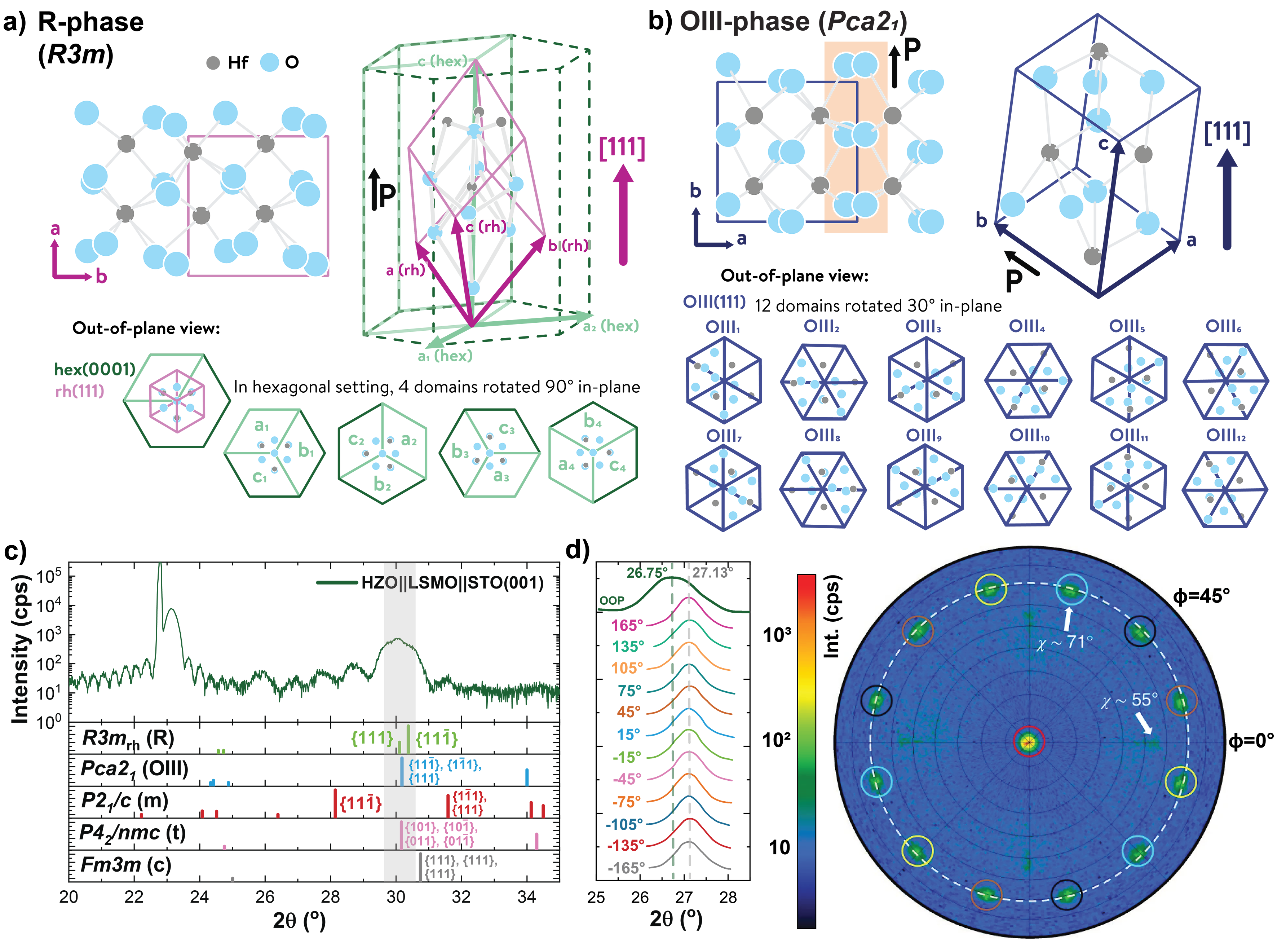}
    \caption{Schematic of a) the \(R3m\) structure and domains in both rhombohedral and hexagonal settings b) the [111]-oriented \(Pca2_1\) OIII-phase and domains, differences between axes length exaggerated for clarity. c) Laboratory source \(\theta/2\theta\) `specular' scan of the HZO \textbar \textbar LSMO\textbar \textbar STO(001) system compared to theoretical predictions for different hafnia polymorphs. d) Pole figure around specular (111) reflection (\(\lambda\)=1.541 Å) and corresponding pole scans compared to the specular (OOP) reflection (\(\lambda\)=1.378 Å) for original R-phase system, adapted from Wei \textit{et al.} \cite{Wei2018AFilms}.}
    \label{fig:INTRO}
\end{figure*}

Ferroelectric materials contain significant promise for non-volatile random access memory, operating faster and more energy-efficiently than existing alternatives \cite{Schenk2020MemoryScientists}. However, this potential has been limited for figurehead ferroelectrics like PbZr$_{1-X}$Ti$_x$O$_3$ and BaTiO$_3$ due to depolarization below 10 nm thicknesses and poor CMOS compatibility \cite{Muller2015FerroelectricProspects,Salahuddin2018TheElectronics,Schenk2020MemoryScientists}. In 2011 a metastable phase of HfO$_2$, a material already well-integrated into CMOS technologies as a high-k dielectric, was reported to exhibit nanoscale ferroelectricity \cite{Boscke2011FerroelectricityFilms}, overcoming both key limiting issues. Ferroelectric behaviour is attributed to a non-centrosymmetric orthorhombic ``OIII-phase", most commonly described by space group $Pca2_1$, Fig.\ref{fig:INTRO}a. This peculiar structure, previously observed in partially-stabilized zirconia (Mg$_{0.1}$Zr$_{0.9}$O$_2$) \cite{Kisi1989CrystalZirconia}, is characterised by a unit cell in which half the oxygen ions contribute to the spontaneous polarization parallel to the \(b\)-axis, while the other half form a non-polar spacer layer \cite{Boscke2011FerroelectricityFilms}.

Despite many varying hypotheses \cite{Lee2020Scale-freeHfO2,Glinchuk2020PossibleFilms, Li2025UnravelingTransitions}, the origin of the uniquely scalable behaviour of ferroelectric hafnia remains unclear. In this context, epitaxial films have been presented as a better platform to investigate the polar nature of the ferroelectric phase when compared to their polycrystalline counterparts \cite{Shimizu2016TheFilm,Wei2018AFilms,Lyu2018RobustFilms} due to reduced structural complexity and a minimization of extrinsic effects. However, reports of a second set of metastable ferroelectric phases with rhombohedral symmetry (R-phase), assigned to space groups \(R3m\) (Fig.\ref{fig:INTRO}b) and \(R3\), for films grown epitaxially on SrTiO$_3$(001)\cite{Wei2018AFilms} and GaN(0001)\cite{Begon-Lours2020StabilizationFilms}, respectively, revealed an additional nuance. In these reports, the R-phases were primarily identified through an elongation of the out-of-plane body diagonal (d$_{111}$) with respect to all (equal) in-plane body diagonals due to large compressive epitaxial strain (Fig. \ref{fig:INTRO}d), with the 1/3 multiplicity of the \{111\} reflections implying a rhombohedral space group. The unlikelihood of a strain-induced tetragonal-rhombohedral or orthorhombic-rhombohedral phase transition suggests that the R-phase is most likely a distortion of the cubic fluorite structure \cite{Wei2018AFilms}, therefore following a different transformation path from the OIII phase reported to evolve from the tetragonal fluorite phase \cite{Boscke2011FerroelectricityFilms,Nentwich2022StructureHfxZr1-xO2, Raeliarijaona2023HafniaFerroelectric}.

The discovery of the R-phase in hafnia-based films was and remains controversial for a number of reasons. Firstly, the formation energies for the R-phases are significantly higher than those of the OIII-phase \cite{Wei2018AFilms, Hu2024PhaseHfO2}. Secondly, the ferroelectric behaviour of R-phase films resembles that of the OIII-phase in both P$_r$ values and scalability, contradicting significantly lower theoretical P$_r$ predictions and leading some to suggest a "rhombohedrally-distorted orthorhombic" phase \cite{Yun2022IntrinsicFilms}. This controversy has remained active due to the difficulty distinguishing the two phases via conventional thin film diffraction approaches, namely; `specular' \straighttheta/2\straighttheta\ measurements, texture (pole figure) and in-plane measurements, and classical reciprocal space mapping (RSMs). As shown in Fig.\ref{fig:INTRO}c, specular studies are insufficient due to the small variation of the lattice parameters relative to a large peak width. The combination of texture and in-plane measurements responsible for the original discovery (Fig\ref{fig:INTRO}d), as well as RSMs \cite{Petraru2024DistinguishingFilms}, can all provide greater insight. However, such measurements have stringent resolution and flux requirements, are rarely performed, and often limited to exploring a small number of reflections. As a result, the current state of epitaxial hafnia literature has become exceedingly confusing. Epitaxial Hf$_{0.5}$Zr$_{0.5}$O$_2$ grown by pulsed laser deposition (PLD) on SrTiO$_3$(001) with La$_{1-x}$Sr$_{x}$MnO$_3$ (LSMO) buffer layers under similar conditions are reported as rhombohedral \cite{Wei2018AFilms, Petraru2024DistinguishingFilms}, orthorhombic \cite{Lyu2018RobustFilms, Shi2023Interface-engineeredFilms} and ``rhombohedrally-distorted orthorhombic" \cite{Yun2022IntrinsicFilms, Kim2025CoerciveEngineering}. Since the existence and orientation of the polarization in ferroelectrics is determined by the crystal symmetry, clarifying the structure of these systems is key to achieve full control of the functional properties.

To remove the remaining ambiguity surrounding the epitaxial R- and OIII-phases requires the expansion of our "field of view" in reciprocal space beyond that provided by the aforementioned diffraction techniques. Here we employ synchrotron based, grazing incidence diffraction with a large area detector to collect extensive and uninterrupted 3D reciprocal space volumes, an approach whose power has been demonstrated recently in other epitaxial systems \cite{Chaney2021TuneableAlloys,Zatterin2024AssessingSuperlattices,Bang2025High-energyFilms}. This technique allows for the reconstruction of arbitrary reciprocal space planes and volumes, providing, for the first time, a complete and undistorted view of reciprocal space for the epitaxial hafnia based systems. In order to clearly compare the two proposed phases, PLD-grown epitaxial hafnia-based films on both SrTiO$_3$ (STO) substrates, reported to be polar rhombohedral \cite{Wei2018AFilms, Petraru2024DistinguishingFilms}, and on Y$_{0.095}$Zr$_{0.905}$O$_2$ (YSZ) substrates, only reported to exhibit the polar orthorhombic phase \cite{Shimizu2016TheFilm} were explored. The ferroelectric properties of both systems have been characterised to confirm structure-functionality links, and their respective transformation pathways have been explored via \emph{in situ} high temperature measurements. Taken together these studies constitute a comprehensive insight into the similarities and key differences of both ferroelectric phases of epitaxial hafnia.

\subsection*{Comparison of 'as grown' diffraction}

\begin{figure*}[!tp]
    \centering
    \includegraphics[width=\linewidth]{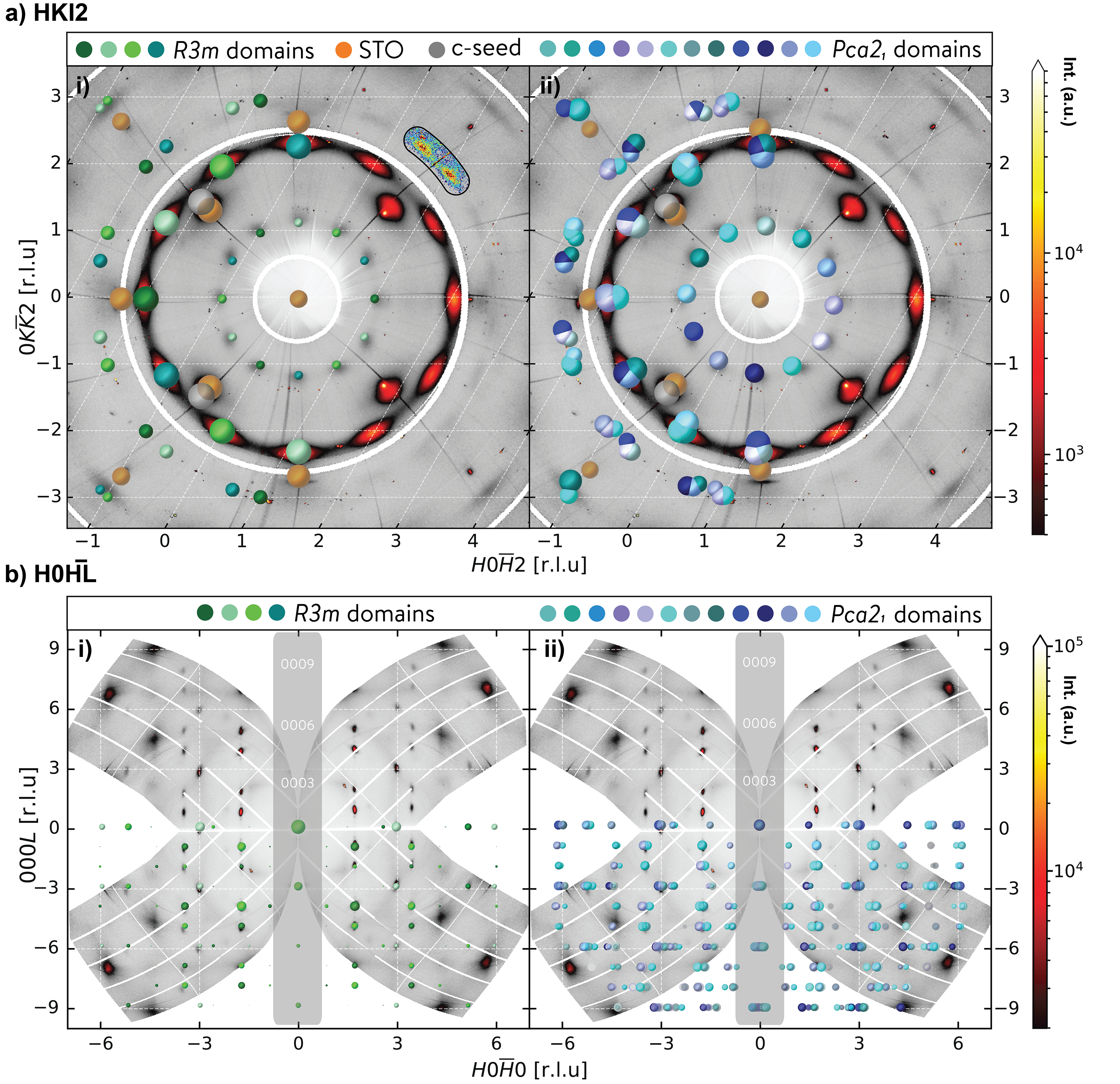}
    \caption{a) In-plane (\(HKI2\)) and b) out-of-plane (\(H0\bar HL\)) reciprocal space reconstructions for a 10 nm YHO$(0001)_{\mathrm{H}}$ film grown on STO(001)/LSMO. Left/right panels show data overlaid by R-/OIII-phase simulations, respectively. Sphere colour represents origin phase and domain, with diameter depicting relative predicted intensity. For a) the simulation has been partially removed to display underlying data, however, the \(Pca2_1\) does not contain this mirror plane, instead the full pattern is produced (not respecting domain origin) by a 180° rotation about the reconstruction normal. In b) measured data (top) is compared to simulation plus data (bottom) related by a 180° rotation about reconstruction normal. In both cases simulations are of a needle-shaped finite reciprocal space slice with depth fading showing proximity to reference plane. The obtained lattice parameters for this R-phase film are \(a_{\mathrm{H}}=b_{\mathrm{H}}=\)6.15 Å, \(c_{\mathrm{H}}\)=8.75 Å (\(\alpha=\beta=90°, \gamma=120°\)), or \(a_{\mathrm{R}}=b_{\mathrm{R}}=c_{\mathrm{R}}=\)4.60 Å (\(\alpha=\beta=\gamma=84.01°\)).  All reconstructions are shown on a ``high dynamic range" mixed logarithmic-linear colour scale. Intensities above a given threshold are mapped onto the logarithmic colorbar shown, with intensities below mapped onto a linear greyscale. The exception being one pair of reflections in (a) highlighted on a separate linear colourscale.}
    \label{fig:SCXRD_R}
\end{figure*}

Starting with the original R-phase system, LSMO-buffered SrTiO$_3$(100) \cite{Wei2018AFilms}. Exemplar reciprocal space reconstructions for a 10 nm Y$_{0.07}$Hf$_{0.93}$O$_{2-x}$ (YHO) film are shown in Fig. \ref{fig:SCXRD_R} and compared with simulated diffraction patterns for both the \(R3m\) and \(Pca2_1\) phases. In-plane reconstructions, i.e. perpendicular to the growth direction, exhibit twelve-fold multiplicity corresponding to four, 90° rotated, epitaxial rhombohedral domains, as previously reported \cite{Wei2018AFilms} and matching the four-fold rotational symmetry of LSMO(001) \cite{Estandia2020Domain-Matching001}. Previous texture measurements of similar films have demonstrated several epitaxial domains, regardless of their ascribed phase \cite{Wei2018AFilms, Song2020EpitaxialFilms}. 

Accounting for domain multiplicity and in-plane mosaicity, Fig.\ref{fig:SCXRD_R}a clearly demonstrates excellent agreement between measured data and simulated \(R3m\) pattern, not only for the major \{\(20\bar22\)\}$_{\mathrm{H}}$ type reflections, but also the subtle domain related peak splitting of the weaker \{\(12\bar32\)\}$_{\mathrm{H}}$ and \{\(13\bar32\)\}$_{\mathrm{H}}$ type reflections. Relative reflection intensities are also well reproduced, consider for example the minor \{\(10\bar12\)\}$_{\mathrm{H}}$ and major \{\(20\bar22\)\}$_{\mathrm{H}}$ type reflections that compose the first and second rings. Further reconstructed planes for the 10 nm YHO film, as well as various stoichiometries, thicknesses and substrate orientations are shown in Fig.S3-4, all showing a consistent picture. By contrast, applying the same procedure for the OIII-phase requires 12 \(Pca2_1\) epitaxial domains to replicate the observed scattering in the $(HKI2)_\mathrm{H}$ plane, in contradiction with the current understanding of domain matching epitaxy for the \(Pca2_1\)\textbar \textbar LSMO(001) system \cite{Estandia2020Domain-Matching001}. Furthermore, no evidence of the characteristic peak splitting inherent to an orthorhombic system, "rhombohedrally distorted" or otherwise, is observed in any of the measured data, an effect that is increasingly obvious for higher order planes (Fig. S2). Nor are relative predicted intensities in agreement with the data, clearly apparent for the inner ring of Fig.\ref{fig:SCXRD_R}a(ii) where the \(Pca2_1\) simulation significantly over-predicts expected intensity. 

\begin{figure*}[!ht]
    \centering
    \includegraphics[width=\linewidth]{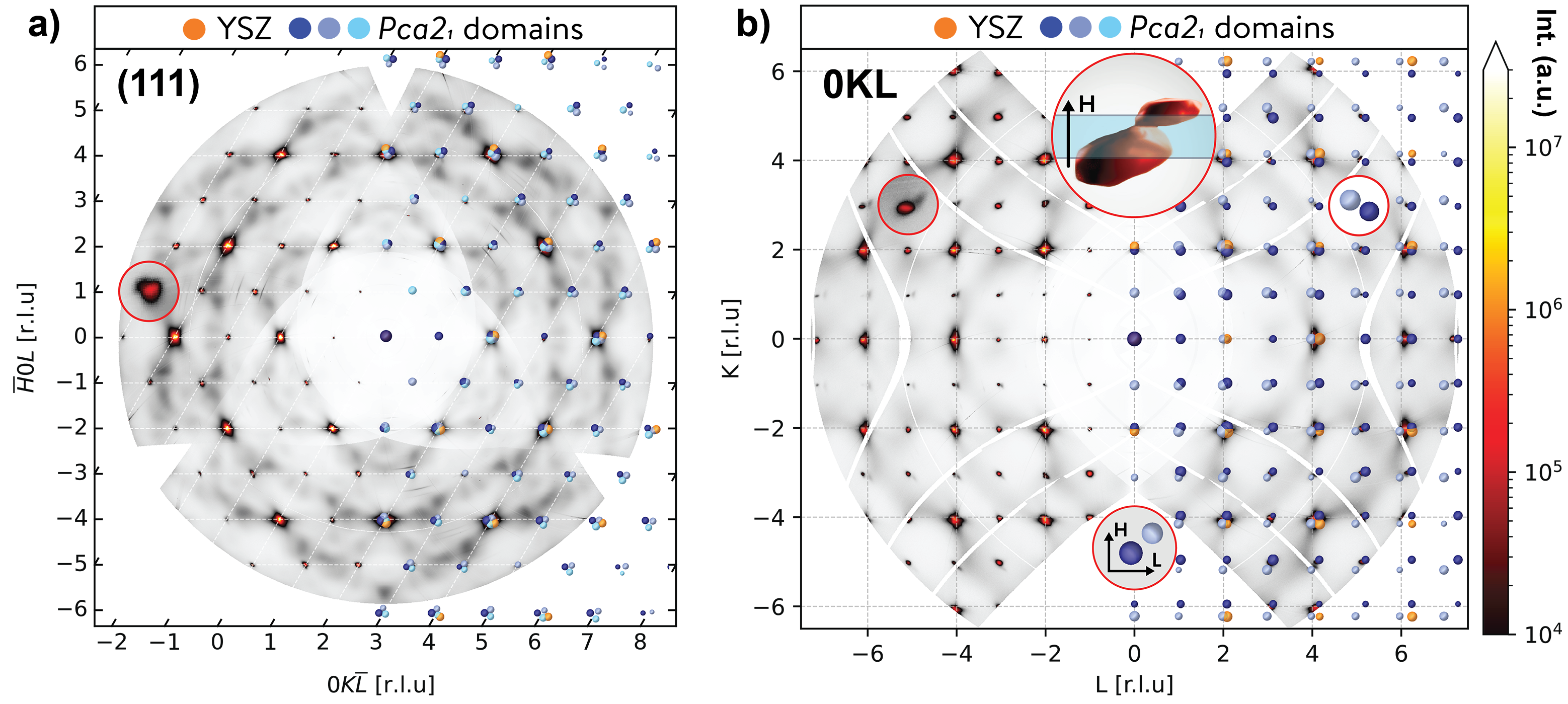}
    \caption{a) (111) and b) \(0KL\) reconstructions for YHO films grown on YSZ(110), note that reconstructions are performed in substrate frame and data has been symmetrized with three-fold and horizontal mirror operations, respectively. Simulated diffraction patterns for \{110\} domains of the OIII-phase (blue) on YSZ(110) (orange) are overlaid in both reconstructions, where the sphere diameter is indicative of relative intensities. 3D peak splitting indicative of OIII domains is highlighted the insets. Note that in b) only two domains are shown since the third maps directly onto the first.}
    \label{fig:SCXRD_OIII}
\end{figure*}

While this agreement is strongly supportive of an R-phase, it is insufficient to consider only in-plane type reconstructions. However, having surveyed a full 3D reciprocal space volume we are able to produce any arbitrary cut, including pure ``out-of-plane" reconstructions like those shown in Fig.\ref{fig:SCXRD_R}b. From which it is immediately clear that we again observe excellent agreement between the observed data and \(R3m\) simulations whereas the \(Pca2_1\) simulations predict both a plethora of additional peaks as well as significant orthorhombic peak splitting, neither of which are observed experimentally.

Atomic structure refinement from grazing incidence data is rarely performed, even from synchrotron data, due to a multitude of complicating factors \cite{May2010QuantifyingFilms, Wadley2013ObtainingRadiation}. However, considering both the \straighttheta/2\straighttheta\ scans (Fig. S1) plus a large number of reconstructions and accounting for four epitaxial domains, we verify all expected rhombohedral reflection conditions accessible within the measured regions of reciprocal space, namely \(-h+k+l=3n\), \(-h+k=3n\) and \(l=3n\). While these conditions are shared for both \(R3\) and \(R3m\), the observed patterns demonstrate closer agreement with the \(R3m\) structure first reported by Wei \emph{et al.} on the same STO substrates \cite{Wei2018AFilms}, with inconsistencies in relative reflection intensities arising when comparing to the \(R3\) structure reported on GaN(0001)-buffered Si substrates \cite{Begon-Lours2020StabilizationFilms, Nukala2020GuidelinesFilms, Ouyang2023StructuralHfO2}. Therefore, considering the excellent agreement between predictions based on a \(R3m\) structure for both the position and intensity of the observed scattered intensity in a large 3D volume of reciprocal space, we conclude that the crystal phase of the hafnia films epitaxially grown on LSMO/STO is indeed rhombohedral, with the most likely space group being \(R3m\). In addition, we can quantify the rhombohedral angle to range from 82.9-84.0° depending on film stoichiometry and thickness (see Table S1 and Fig. S3), a larger deviation from 90° than previously reported \cite{Wei2018AFilms,Yun2022IntrinsicFilms}.

\begin{figure*}[!ht]
    \centering
    \includegraphics[width=\linewidth]{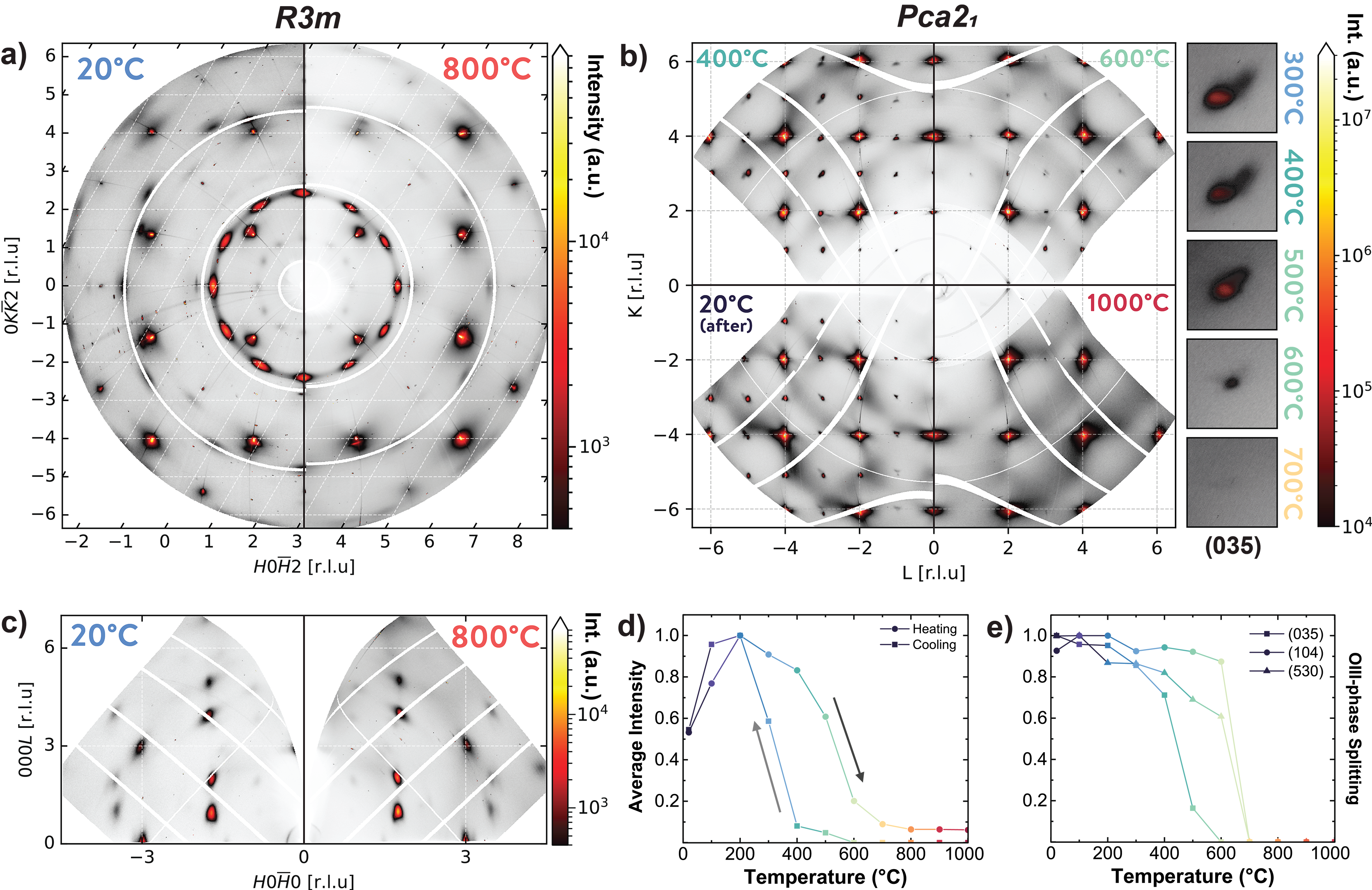}
    \caption{a) \(HKI2\) and c) \(H0\bar HL\) reconstructions for R-phase films at 20°C and 800°C (growth temperature). b) Comparison of the \(0KL\) reconstructions for OIII(101) films from 400°C to 700°C. Right-hand panel shows evolution of the (035) peaks with temperature.  d) Average normalized OIII-phase intensity with temperature upon heating and cooling. e) OIII-splitting for select peaks as a function of temperature.}
    \label{fig:SCXRD_PT}
\end{figure*}

Before starting our inspection of the YHO\textbar \textbar YSZ system, we also report a [001]-oriented cubic (\(Fm\bar3m\)) seed layer rotated 45° in-plane with respect to the SrTiO$_3$(001) substrate, observed in both LSMO-buffered, see grey spheres Fig.\ref{fig:SCXRD_R}a, and films grown directly on STO, Fig. S8. The seed layer lattice parameters coincide with those observed by scanning transmission electron microscopy (STEM) in similar R-phase films \cite{Wei2018AFilms} but attributed to a tetragonal interfacial layer. Given that when observed along the [100] zone-axis of a STO(001) substrate, a [001]-oriented cubic layer rotated 45° in-plane and a [010]-oriented tetragonal layer parallel to the substrate are difficult to distinguish via STEM (see Fig. S6) it seems likely that this interfacial layer was actually [001]-oriented cubic (\(Fm\bar3m\)). This observation that R-phase growth initiates with the parent hafnia polymorph supports the hypothesis that epitaxial strain can only induce the rhombohedral distortion from the cubic phase. Follow-up atomic force microscopy (AFM) characterisation (Fig. S7) suggests the resulting large compressive strain is partially alleviated by the formation of grain boundaries as we observe the R-phase films grows in epitaxial columnar grains, in contrast to the YHO \textbar \textbar YSZ system, which exhibits atomically flat surfaces. The formation of grains may also help stabilise the high-energy R-phase by increasing the surface energy as has been shown previously for epitaxial R-phase ZrO$_2$. \cite{ElBoutaybi2022StabilizationEnergy, Wang2026GrainFerroelectrics}

To make a direct comparison we now consider YHO films grown directly on YSZ(110) under identical conditions to the R-phase films, with the resulting reciprocal space reconstructions shown in Fig. \ref{fig:SCXRD_OIII}. Unlike for the R-phase system, a buffer is not explicitly required for phase stabilisation and using a directly deposited system simplifies interpretation. It is immediately evident that, unlike for the previous system, we observe an excellent match between measured scattering and simulated OIII patterns. Considering the (111)$_{\mathrm{O}}$ type reconstruction shown in Fig.\ref{fig:SCXRD_OIII}a, we require only 3 epitaxial domains, and the characteristic triangular peak splitting expected from OIII lattice parameter mismatch is clearly resolvable for high-Q reflections. Such splitting is particularly evident in the (0KL)$_{\mathrm{O}}$ reconstruction, Fig.\ref{fig:SCXRD_OIII}b, where some domain reflections lie above or below the plane as a result of the reconstruction being in the substrate frame and can be clearly resolved by 3D isosurface reconstructions. Accounting for these factors, predicted and observed intensities agree well throughout the 3D reciprocal space volume collected and we can therefore conclude that, as expected, the YHO \textbar \textbar YSZ(110) system is well described by three epitaxial domains of the standard \(Pca2_1\) ``OIII" structure and that, by sampling large portions reciprocal space the two ferroelectric phases of epitaxial hafnia are in fact easily distinguishable. 

\subsection*{High temperature phase transitions}
To gain greater insight into the phase stability and transformation pathways of the two structures we performed \textit{in-situ} grazing incidence diffraction measurements up to 1000°C in a nitrogen environment employing a purpose built furnace to minimize background contributions, see Fig. S12. 

For the R-phase system, shown in Fig. \ref{fig:SCXRD_PT}a, we observe no phase transition up to the growth temperature (800°C) with both the cubic seed and R-phase reflections remaining largely unchanged. Thus implying that both layers grow directly in their respective structures and there is no need for a stable-to-metastable phase transition on cooling. Above 800°C, thermal degradation at the STO/LSMO interface driven by Sr diffusion makes measurements uninterpretable and we are thus unable to access any possible R-phase to cubic transition. Unfortunately, the requirement for LSMO buffer is well known \cite{Schroeder2019FerroelectricityDevices,Wei2018AFilms,Estandia2021CriticalFilms,Shi2023Interface-engineeredFilms} and, in its absence, our films exhibit lower crystallinity and a mixture of both epitaxial and non-epitaxial rhombohedral and monoclinic phases, albeit the cubic seed layer is preserved (see Fig. S8). 

By contrast, the OIII-phase YHO\textbar \textbar YSZ(110) films show a clear phase transition with increasing temperature, as seen in Fig.\ref{fig:SCXRD_PT}b. An end state is reached above $\sim$700°C, with no remaining film intensity observable, implying a transition to the high temperature cubic phase and a collapse of all scattered intensity onto substrate reflections positions. Evidently, this cannot be confirmed directly due to the overwhelming intensity of the substrate reflections, however, no indication of low-Q scattering or diffraction rings are observed and the film is thus assumed to remain crystalline at the maximum temperature. Studying the OIII peak evolution more carefully (Fig.\ref{fig:SCXRD_PT}c) reveals that at $\sim$450°C \(b=c\) while \(a\neq c\) until $\sim$700°C, implying the presence of a tetragonal intermediate in the 450-700°C range. However, due to the presence of domains, it is not possible to directly observe the systematic absences of the expected tetragonal intermediate. Both the Curie temperature (T$_c$) and the step-wise nature of the phase transition coincide with previous findings from Shimizu \emph{et al.} on YHO/YSZ(110) epitaxial films \cite{Shimizu2016TheFilm}. 

In addition, we demonstrate that this phase transition is reversible and comparison of the normalised OIII-phase intensity upon heating and cooling reveals hysteretic behaviour indicative of a first-order phase transition, as expected for the martensitic phase transition that takes place in bulk hafnia and zirconia above 1000 °C \cite{Wolten1963DiffusionlessHafnia}. Confirming that, in contrast to R-phase films, OIII-phase films initially grow in the cubic phase before a transition to OIII through a tetragonal intermediate upon cooling. Given this, one might expect that the OIII phase stability be highly sensitive to cooling rate, however, post-growth quenching from 1000°C (\(\sim\) 300°C/min) also leads to recovery of the OIII-phase (see Fig. S10), implying cooling rate is not crucial to OIII phase stability.

\subsection*{Comparing ferroelectric performance}
\begin{figure*}[!ht]
    \centering
    \includegraphics[width=\linewidth]{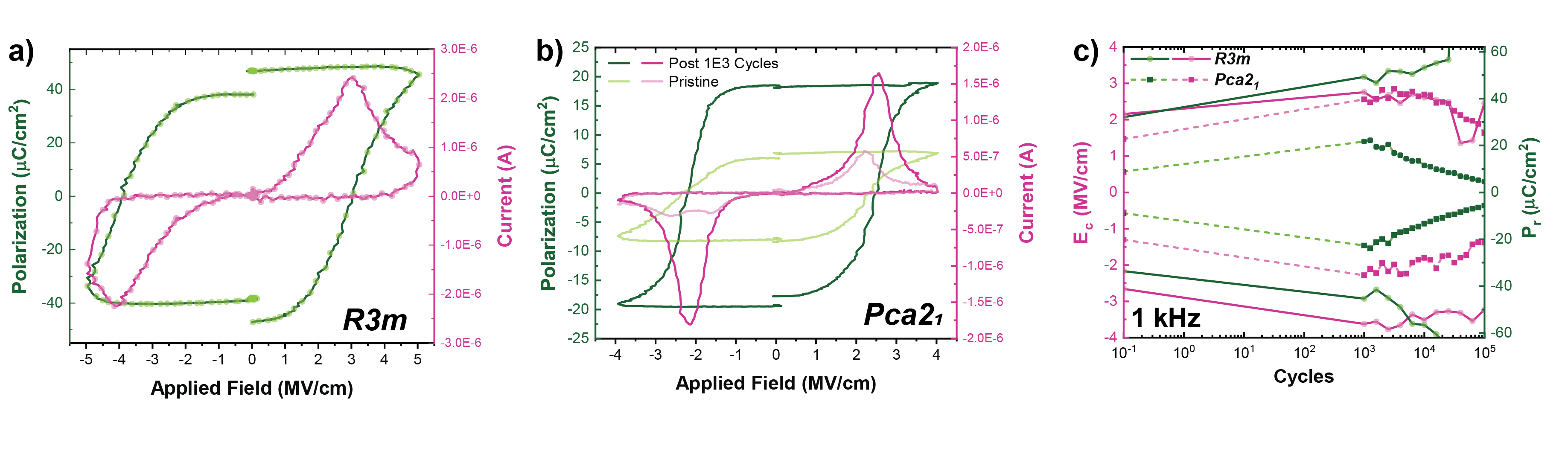}
    \caption{P-E loops and and I-E loops for a) R-phase and b) OIII-phase (right) Hf$_{0.25}$Zr$_{0.75}$O$_2$(111) 10 nm films. c) Evolution of the remanent polarization (P$_r$) and coercivity (E$_c$) upon cycling at 1 kHz. Solid/dashed lines with opaque/transparent symbols, correspond to R-phase and OIII-phase, respectively.}
    \label{fig:EComp}
\end{figure*}

As epitaxial films, both systems have a well-defined polar axis in the as-grown state. It is, therefore, surprising that the biggest difference between the electrical response of both films (Fig. \ref{fig:EComp}) is the absence of a wake-up effect for the R-phase system, which exhibits a remanent polarisation $(P_r)$ of 40 \(\mu\) C/cm$^2$ as-grown. In contrast, the equivalent OIII-phase system reaches a maximum of 20 \(\mu\)C/cm$^2$ only after 1000 cycles of wake-up, with an almost twofold increase in P$_r$ compared to as-grown. Note that electrical measurements were performed on HZO rather than YHO films as the former exhibits more pronounced ferroelectric characteristics for both equivalent systems. For the OIII-phase, an ITO-buffered (111)$_\mathrm{O}$ film was employed to allow direct comparison to the (111)$_\mathrm{R}$ film, with the ITO buffer acting as the bottom electrode. Importantly, at ambient conditions, qualitatively comparable diffraction is confirmed between the R- and OIII-phase pairs used for electrical and diffraction studies, see Fig. S1.

That the P$_r$ of the OIII(111) system is below the theoretical value is expected given the polar axis is not purely out-of-plane and only a component of the true polarization is measured. This also means the applied field and polar axis are non-parallel and the observed wake-up process could be explained by a (ferroelastic) reorientation of \{111\} epitaxial domains under applied field, where the domains with the largest proportion of out-of-plane polar axis are favoured, leading to the observed increase in P$_r$. Similarly, the subsequent rapid fatigue may result from defect formation associated with the application of an electric field not parallel to the axis of switching polarization. 

For the R-phase (111)-system, the P$_r$ consistently exceeds theoretical expectations \cite{Wei2018AFilms}. Given that the P$_r$ is expected to increase with increasing d$_{111}$ \cite{Wei2018AFilms}, this can be partially accounted for by considering that the presently measured rhombohedral distortion is larger than those previously reported. In addition, we find a large substoichiometric deviation in oxygen content away from HfO$_2$ in R-phase films, which is also expected to influence theoretical P$_r$ values. Via \textit{ex-situ} x-ray photoemission spectroscopy (XPS) we measure significantly lower oxygen stoichiometries (Hf$_{0.5}$Zr$_{0.5}$O$_{1.72}$) for R-phase films, compared to the OIII-phase films grown under comparable conditions, (Hf$_{0.5}$Zr$_{0.5}$O$_{1.86}$), see Fig. S11. This is in agreement with previous reports of R-phase films requiring significantly greater oxygen substoichiometries for phase stability when compared to OIII-phase films\cite{Kaiser2023CrystalOxide, Zhang2024PhaseFatigue}.

Aside from contrasting behaviour upon cycling, there are clear differences in the shape and coercivity of the measured current loops. The R-phase shows; imprint indicating the presence of a negative internal bias, large coercive fields, and switching peaks with a much wider distribution. The sharper switching observed for the OIII-phase is more akin to the traditional ferroelectrics PbTiO$_3$ and BaTiO$_3$, while the broad switching distribution observed for the R-phase is indicative of switching inhomogeneities across the sample, possibly related to the higher oxygen substoichiometry of these films. Meanwhile, the OIII-phase exhibits no obvious imprint effect, lower coercivity and narrower switching peaks (after wake-up). The difference in imprint across both devices is likely related to the differing bottom electrode resulting in different Schottky barriers and oxygen diffusion behaviour. Finally, comparison of the fatigue measurements performed at 1 kHz show the endurance of both types of devices is limited to a similar number of cycles but through different processes: filamentary breakdown and ferroelectric fatigue for the R- and OIII-phases, respectively.

\subsection*{Discussion}
In this paper, we conclusively demonstrate the ease with which the ferroelectric rhombohedral (\(R3m\)) and orthorhombic (\(Pca2_1\)) phases of hafnia can be distinguished when moving beyond classic thin film diffraction techniques. We resolve the long standing phase ambiguity and show that epitaxial hafnium-based ferroelectric films on LSMO\textbar \textbar STO are rhombohedral, most likely \(R3m\), which undoubtedly places the ferroelectric polarization along the most favorable [111] (out-of-plane) direction (parallel to the applied field in capacitive and tunnel junction devices). Meanwhile, films deposited under identical conditions on YSZ are well described by the classical OIII structure, demonstrating that it is possible to select the desired ferroelectric phase through substrate choice, where a large compressive strain is key to the stabilising the R-phase over the otherwise lower formation energy OIII-phase. Finally, state-of-the-art \emph{in situ} temperature-dependent measurements demonstrated R-phase phase stability up to 800°C, showing its robustness and implying that the rhombohedral structure forms directly during growth along with an initial cubic seed layer. This contrasts with equivalent measurements on YHO(110)\textbar \textbar YSZ(110) showing a two-step, reversible first-order phase transition from the OIII-phase to a cubic phase, through a tetragonal intermediate, over the range of 300-700°C.

Having decisively disentangled the two phases, we were able to compare electrical characterization of both, showing clear differences in switching kinetics, where R-phase films present broader switching peaks, no wake-up effect and endurance limited by filamentary breakdown, while OIII-phase films show substantially sharper switching peaks, a clear wake-up effect and are limited by ferroelectric fatigue. Currently the electrical performance of epitaxial R-phase films has been superior \cite{Shimizu2016TheFilm,Yun2022IntrinsicFilms}. However, whether inadvertently or not, significantly more effort has been devoted to optimization of epitaxial R-phase devices. From our results, it seems plausible that epitaxial OIII-phase films could outperform R-phase films regarding both P$_r$ and endurance if exclusively (010)-oriented OIII films (polar axis fully out-of-plane) could be achieved, assuming that the wake-up and fatigue effects observed here for (111) films are, indeed, related to an off-axis applied field. However, given the similar length of the \(a\) and \(b\) axes in the OIII structure, epitaxial strain alone may not produce pure (010) films \cite{Katayama2016OrientationFilms}. Resolving this should be of great interest to the ferroelectric hafnia community, particularly considering that recently OIII-phase films of Hf$_{0.5}$Zr$_{0.5}$O$_2$ were grown epitaxially by ALD \cite{Cho2024AtomicFilms}.

We believe that the systematic and extensive studies presented in this work have removed the long standing ambiguity surrounding the phase-nature of epitaxial ferroelectric hafnia-based films, highlighting the power of large 3D reciprocal space surveys to provide a complete picture of the epitaxial system of interest and allow for the establishment of accurate structure-functionality links. It is our hope that the utilisation of such techniques becomes increasingly common for exploring not only epitaxial hafnia based systems but any ferroelectric, or otherwise exciting, epitaxial system, for which we have been commonly deprived of a wealth of information by restricting studies to only limited regions of reciprocal space.

\subsection*{Experimental Methods}
\subsubsection*{Sample Growth}
Epitaxial hafnia films are synthesized by pulsed laser deposition (PLD) on single crystal substrates: SrTiO$_3$ (STO) and yttria-stabilized zirconia (YSZ, 9.5 mol\% Y$_2$O$_3$), using a 248 nm excimer laser for ablation. For films on STO, a (conducting) 27 nm buffer layer of La$_{0.67}$Sr$_{0.33}$MnO$_3$ (LSMO) is included, while on YSZ, a 20 nm indium tin oxide (ITO, In$_2$O$_3$:SnO$_2$, 90:10 wt\%) buffer layer is employed as a bottom electrode. Both LSMO and ITO targets were purchased from PI-KEM, while hafnia targets of different compositions (HfO$_2$, Hf$_{0.50}$Zr$_{0.50}$O$_2$, Hf$_{0.25}$Zr$_{0.75}$O$_2$, Y$_{0.07}$Hf$_{0.93}$O$_{2-x}$) were synthesized by a solid-state reaction at 1400°C using HfO$_2$ (99\% purity), ZrO$_2$ (99.5\% purity) and Y$_2$O$_3$ (99.99\% purity) powders. LSMO layers of ~30 nm are deposited on STO substrates heated to 775°C using a laser fluence of 1.8 J/cm$^2$ and 1 Hz laser frequency under a 0.15 mbar oxygen atmosphere. ITO layers of ~20 nm are deposited on YSZ substrates heated to 650°C under a 0.05 mbar oxygen atmosphere by interval deposition, consisting of 50 pulse bursts at a laser fluence of 1.5 J/cm$^2$ and 20 Hz frequency with 2 min intervals. In all cases hafnia/zirconia layers are deposited under the same conditions; substrate temperature of 800°C, laser fluence of 2 J/cm$^2$ and 2 Hz frequency under a 0.05 mbar oxygen atmosphere. After deposition, all samples were cooled to room temperature at a rate of 5°C/min in an oxygen pressure of 300 mbar.

\subsubsection*{X-ray Diffraction}
Specular \(\theta/2\theta\) measurements are performed using a Cu K\(\alpha\) lab source (\(\lambda\) = 1.54 Å) on a Panalytical X'Pert Pro diffractometer in line focus.

Grazing incidence x-ray diffraction measurements were performed at the ID28 beamline \cite{Girard2019AESRF}, European Synchrotron Radiation Facility (ESRF) using a Pilatus3 X 1M detector operated in shutterless mode with continuous rotation of the sample. Unless otherwise specified, measurements were performed with an incident wavelength of 0.7839 Å, low energy threshold of $12\;\mathrm{keV}$, integrating over an angular range of 0.25° for a time of $1\;\mathrm{s}$, at $45\degree$ rotation around incident beam and with detector positions of $19\degree$ and $48\degree$ to maximise reciprocal space coverage. The effective grazing angle was optimised for scattered hafnia signal in each sample and ranged from 0.1° to 0.6°. \emph{In situ} temperature-dependent measurements were performed using a custom furnace design under a nitrogen atmosphere, further details in SI. Effective grazing angle and sample height were again optimised for hafnia signal at each temperature. Raw frames are processed using the Crysalis Pro suite \cite{2014CrysAlisPRO} to obtain a UB matrix, using which both 2D and 3D reciprocal space reconstructions are produced using in-house software developed at ID28. 3D reconstructions are finally rendered in Blender. Expected diffraction signal for given reciprocal space reconstructions are produced from literature structures (adjusted for the experimental lattice parameters and stoichiometry) Wei et al. \cite{Wei2018AFilms} and Kisi \textit{et al.} \cite{Kisi1989CrystalZirconia} for \(R3m\) and \(Pca2_1\), respectively, using the 3D option within the CrystalMaker SingleCrystal software \cite{2014CrystalMaker} and employing needle-like depth-fading.

\subsubsection*{Atomic force microscopy}
Atomic force microscopy (AFM) is performed using BudgetSensors Tap300AI-G (300 kHz, 40 N/m) tips on a Bruker Dimension Icon in tapping mode. 

\subsubsection*{X-ray photoelectron spectroscopy}
X-ray photoelectron spectroscopy (XPS) is performed using an Al K\(\alpha\) source operated at 34 mA and 10 kV with a 54.7° incidence angle on the sample.

\subsubsection*{Ferrolectric Characterisation}
Circular Ti(4 nm)/Au(40 nm) top electrodes of various sizes (10-50 µm diameter) are patterned by photolithography and deposited by e-beam evaporation in order to enable electrical characterization. Ferroelectric measurements are performed using an aixACCT TF Analyser 2000, where P-E loops are collected from pristine films using 1 kHz PUND measurements \cite{Scott1988SwitchingMemories}, while cyclability is tested using fatigue measurements (100 kHz, 1 kHz PUND).

\subsection*{Data Availability}
Data collected at the European Synchrotron Radiation Facility as part of proposal HC-5778 is available at the following DOI: https://doi.org/10.15151/ESRF-ES-1560220794. 

Further data supporting this study is available in DataverseNL at INSERTLINKHERE.

\subsection*{References}
\singlespacing
\renewcommand*{\bibfont}{\small}
\printbibliography[heading=none]

@article{Schenk2020MemoryScientists,
    title = {{Memory technology—a primer for material scientists}},
    year = {2020},
    journal = {Reports on Progress in Physics},
    author = {Schenk, T. and Pe{\v{s}}i{\'{c}}, M. and Slesazeck, S. and Schroeder, U. and Mikolajick, T.},
    number = {8},
    month = {6},
    pages = {086501},
    volume = {83},
    publisher = {IOP Publishing},
    url = {https://iopscience.iop.org/article/10.1088/1361-6633/ab8f86 https://iopscience.iop.org/article/10.1088/1361-6633/ab8f86/meta},
    doi = {10.1088/1361-6633/AB8F86},
    issn = {0034-4885},
    pmid = {32357345}
}

@article{Muller2015FerroelectricProspects,
    title = {{Ferroelectric Hafnium Oxide Based Materials and Devices: Assessment of Current Status and Future Prospects}},
    year = {2015},
    journal = {ECS Journal of Solid State Science and Technology},
    author = {M{\"{u}}ller, J and Polakowski, P and Mueller, S and Mikolajick, T},
    number = {5},
    pages = {N30-N35},
    volume = {4},
    publisher = {The Electrochemical Society},
    url = {http://jss.ecsdl.org/content/4/5/N30.full.pdf},
    doi = {10.1149/2.0081505jss},
    issn = {2162-8769}
}

@article{Salahuddin2018TheElectronics,
    title = {{The era of hyper-scaling in electronics}},
    year = {2018},
    journal = {Nature Electronics},
    author = {Salahuddin, Sayeef and Ni, Kai and Datta, Suman},
    number = {8},
    pages = {442--450},
    volume = {1},
    url = {https://doi.org/10.1038/s41928-018-0117-x},
    doi = {10.1038/s41928-018-0117-x},
    issn = {2520-1131}
}

@article{Boscke2011FerroelectricityFilms,
    title = {{Ferroelectricity in hafnium oxide thin films}},
    year = {2011},
    journal = {Applied Physics Letters},
    author = {B{\"{o}}scke, T S and M{\"{u}}ller, J and Br{\"{a}}uhaus, D and Schr{\"{o}}der, U and B{\"{o}}ttger, U},
    number = {10},
    pages = {102903},
    volume = {99},
    publisher = {AIP Publishing},
    url = {https://aip-scitation-org.proxy-ub.rug.nl/doi/full/10.1063/1.3634052},
    doi = {10.1063/1.3634052},
    issn = {0003-6951}
}

@article{Kisi1989CrystalZirconia,
    title = {{Crystal Structure of Orthorhombic Zirconia in Partially Stabilized Zirconia}},
    year = {1989},
    journal = {Journal of the American Ceramic Society},
    author = {Kisi, Erich H and Howard, Christopher J and Hill, Roderick J},
    number = {9},
    pages = {1757--1760},
    volume = {72},
    publisher = {Wiley},
    url = {https://dx.doi.org/10.1111/j.1151-2916.1989.tb06322.x},
    doi = {10.1111/j.1151-2916.1989.tb06322.x},
    issn = {0002-7820}
}

@article{Lee2020Scale-freeHfO2,
    title = {{Scale-free ferroelectricity induced by flat phonon bands in HfO2}},
    year = {2020},
    journal = {Science},
    author = {Lee, Hyun-Jae and Lee, Minseong and Lee, Kyoungjun and Jo, Jinhyeong and Yang, Hyemi and Kim, Yungyeom and Chae, Seung Chul and Waghmare, Umesh and Lee, Jun Hee},
    number = {6509},
    pages = {1343--1347},
    volume = {369},
    url = {https://www.science.org/doi/abs/10.1126/science.aba0067},
    doi = {doi:10.1126/science.aba0067}
}

@article{Glinchuk2020PossibleFilms,
    title = {{Possible electrochemical origin of ferroelectricity in HfO2 thin films}},
    year = {2020},
    journal = {Journal of Alloys and Compounds},
    author = {Glinchuk, Maya D. and Morozovska, Anna N. and Lukowiak, Anna and Strck, Wiesław and Silibin, Maxim V. and Karpinsky, Dmitry V. and Kim, Yunseok and Kalinin, Sergei V.},
    month = {7},
    pages = {153628},
    volume = {830},
    publisher = {Elsevier},
    doi = {10.1016/J.JALLCOM.2019.153628},
    issn = {0925-8388},
    arxivId = {1811.09787}
}

@article{Li2025UnravelingTransitions,
    title = {{Unraveling the origins of ferroelectricity in doped hafnia through carrier-mediated phase transitions}},
    year = {2025},
    journal = {npj Computational Materials},
    author = {Li, Gang and Yan, Shaoan and Liu, Yulin and Zhang, Wanli and Xiao, Yongguang and Yang, Qiong and Tang, Minghua and Li, Jiangyu and Long, Zhilin},
    number = {1},
    month = {12},
    pages = {1--11},
    volume = {11},
    publisher = {Nature Research},
    url = {https://www-nature-com.proxy-ub.rug.nl/articles/s41524-025-01515-7},
    doi = {10.1038/S41524-025-01515-7;SUBJMETA=119,301,544,639,996;KWRD=FERROELECTRICS+AND+MULTIFERROICS,SURFACES},
    issn = {20573960}
}

@article{Shimizu2016TheFilm,
    title = {{The demonstration of significant ferroelectricity in epitaxial Y-doped HfO2 film}},
    year = {2016},
    journal = {Scientific Reports 2016 6:1},
    author = {Shimizu, Takao and Katayama, Kiliha and Kiguchi, Takanori and Akama, Akihiro and Konno, Toyohiko J. and Sakata, Osami and Funakubo, Hiroshi},
    number = {1},
    month = {9},
    pages = {1--8},
    volume = {6},
    publisher = {Nature Publishing Group},
    url = {https://www.nature.com/articles/srep32931},
    doi = {10.1038/srep32931},
    issn = {2045-2322}
}

@article{Wei2018AFilms,
    title = {{A rhombohedral ferroelectric phase in epitaxially strained Hf0.5Zr0.5O2 thin films}},
    year = {2018},
    journal = {Nature Materials},
    author = {Wei, Yingfen and Nukala, Pavan and Salverda, Mart and Matzen, Sylvia and Zhao, Hong Jian and Momand, Jamo and Everhardt, Arnoud S and Agnus, Guillaume and Blake, Graeme R and Lecoeur, Philippe and Kooi, Bart J and {\'{I}}{\~{n}}iguez, Jorge and Dkhil, Brahim and Noheda, Beatriz},
    number = {12},
    pages = {1095--1100},
    volume = {17},
    publisher = {Nature Publishing Group},
    url = {https://www-nature-com.proxy-ub.rug.nl/articles/s41563-018-0196-0},
    doi = {doi:10.1038/s41563-018-0196-0},
    issn = {1476-46601476-4660},
    language = {En}
}

@article{Lyu2018RobustFilms,
    title = {{Robust ferroelectricity in epitaxial Hf1/2Zr1/2O2 thin films}},
    year = {2018},
    journal = {Applied Physics Letters},
    author = {Lyu, J. and Fina, I. and Solanas, R. and Fontcuberta, J. and S{\'{a}}nchez, F.},
    number = {8},
    month = {8},
    pages = {82902},
    volume = {113},
    publisher = {American Institute of Physics Inc.},
    url = {/aip/apl/article/113/8/082902/37051/Robust-ferroelectricity-in-epitaxial-Hf1-2Zr1-2O2},
    doi = {10.1063/1.5041715/37051},
    issn = {00036951}
}

@article{Begon-Lours2020StabilizationFilms,
    title = {{Stabilization of phase-pure rhombohedral HfZrO4 in pulsed laser deposited thin films}},
    year = {2020},
    journal = {Physical Review Materials},
    author = {B{\'{e}}gon-Lours, Laura and Mulder, Martijn and Nukala, Pavan and De Graaf, Sytze and Birkh{\"{o}}lzer, Yorick A and Kooi, Bart and Noheda, Beatriz and Koster, Gertjan and Rijnders, Guus},
    number = {4},
    volume = {4},
    publisher = {American Physical Society (APS)},
    issn = {2475-9953}
}

@article{Nentwich2022StructureHfxZr1-xO2,
    title = {{Structure relations in the family of the solid solution HfxZr1-xO2}},
    year = {2022},
    journal = {Zeitschrift fur Kristallographie - Crystalline Materials},
    author = {Nentwich, Melanie},
    number = {4-5},
    month = {5},
    pages = {141--157},
    volume = {237},
    publisher = {De Gruyter Open Ltd},
    url = {https://www.degruyterbrill.com/document/doi/10.1515/zkri-2021-2066/html},
    doi = {10.1515/zkri-2021-2066},
    issn = {21967105}
}

@article{Raeliarijaona2023HafniaFerroelectric,
    title = {{Hafnia is a proper ferroelectric}},
    year = {2023},
    journal = {Physical Review B},
    author = {Raeliarijaona, Aldo and Cohen, R. E.},
    number = {9},
    month = {9},
    pages = {094109},
    volume = {108},
    publisher = {American Physical Society},
    url = {https://journals.aps.org/prb/abstract/10.1103/PhysRevB.108.094109},
    doi = {10.1103/PhysRevB.108.094109},
    issn = {24699969},
    arxivId = {2305.19446}
}

@article{Hu2024PhaseHfO2,
    title = {{Phase Stability and Phase Transition Pathways in the Rhombohedral Phase of HfO2}},
    year = {2024},
    journal = {Journal of Physical Chemistry Letters},
    author = {Hu, Qi and Lv, Shuning and Xue, Chuang and Tsai, Hsiaoyi and Cao, Tengfei and Xu, Zhongfei and Teobaldi, Gilberto and Liu, Li Min},
    month = {9},
    pages = {9319--9325},
    publisher = {American Chemical Society},
    url = {https://pubs.acs.org/doi/full/10.1021/acs.jpclett.4c02019},
    doi = {10.1021/ACS.JPCLETT.4C02019/ASSET/IMAGES/LARGE/JZ4C02019{\_}0004.JPEG},
    issn = {19487185}
}

@article{Yun2022IntrinsicFilms,
    title = {{Intrinsic ferroelectricity in Y-doped HfO2 thin films}},
    year = {2022},
    journal = {Nature Materials},
    author = {Yun, Yu and Buragohain, Pratyush and Li, Ming and Ahmadi, Zahra and Zhang, Yizhi and Li, Xin and Wang, Haohan and Li, Jing and Lu, Ping and Tao, Lingling and Wang, Haiyan and Shield, Jeffrey E and Tsymbal, Evgeny Y and Gruverman, Alexei and Xu, Xiaoshan},
    number = {8},
    pages = {903--909},
    volume = {21},
    publisher = {Springer Science and Business Media LLC},
    doi = {10.1038/s41563-022-01282-6},
    issn = {1476-1122}
}

@article{Petraru2024DistinguishingFilms,
    title = {{Distinguishing the Rhombohedral Phase from Orthorhombic Phases in Epitaxial Doped HfO2 Ferroelectric Films}},
    year = {2024},
    journal = {ACS Applied Materials and Interfaces},
    author = {Petraru, Adrian and Gronenberg, Ole and Sch{\"{u}}rmann, Ulrich and Kienle, Lorenz and Droopad, Ravi and Kohlstedt, Hermann},
    number = {32},
    month = {8},
    pages = {42534--42545},
    volume = {16},
    publisher = {American Chemical Society},
    url = {/doi/pdf/10.1021/acsami.4c10423},
    doi = {10.1021/ACSAMI.4C10423/ASSET/IMAGES/LARGE/AM4C10423{\_}0011.JPEG},
    issn = {19448252},
    pmid = {39102275}
}

@article{Shi2023Interface-engineeredFilms,
    title = {{Interface-engineered ferroelectricity of epitaxial Hf0.5Zr0.5O2 thin films}},
    year = {2023},
    journal = {Nature Communications 2023 14:1},
    author = {Shi, Shu and Xi, Haolong and Cao, Tengfei and Lin, Weinan and Liu, Zhongran and Niu, Jiangzhen and Lan, Da and Zhou, Chenghang and Cao, Jing and Su, Hanxin and Zhao, Tieyang and Yang, Ping and Zhu, Yao and Yan, Xiaobing and Tsymbal, Evgeny Y. and Tian, He and Chen, Jingsheng},
    number = {1},
    month = {3},
    pages = {1--8},
    volume = {14},
    publisher = {Nature Publishing Group},
    url = {https://www.nature.com/articles/s41467-023-37560-3},
    doi = {10.1038/s41467-023-37560-3},
    issn = {2041-1723},
    pmid = {36997572}
}

@article{Kim2025CoerciveEngineering,
    title = {{Coercive Field Control in Epitaxial Ferroelectric Hf0.5Zr0.5O2 Thin Films by Nanostructure Engineering}},
    year = {2025},
    journal = {ACS Applied Materials and Interfaces},
    author = {Kim, Ji Soo and Strkalj, Nives and Silva, Alexandre and Lenzi, Veniero and Marques, Luis and Hill, Megan O. and Yuan, Ziyi and Liu, Yi Xuan and Becker, Maximilian T. and Fairclough, Simon M. and Ducati, Caterina and Zhang, Yizhi and Shen, Jianan and Hu, Zedong and Dou, Hongyi and Wang, Haiyan and Silva, José P.B. and MacManus-Driscoll, Judith L.},
    month = {4},
    pages = {25442--2540},
    volume = {17},
    publisher = {American Chemical Society},
    doi = {10.1021/ACSAMI.4C21787/ASSET/IMAGES/LARGE/AM4C21787{\_}0004.JPEG},
    issn = {19448252}
}

@article{Chaney2021TuneableAlloys,
    title = {{Tuneable correlated disorder in alloys}},
    year = {2021},
    journal = {Physical Review Materials},
    author = {Chaney, D. and Castellano, A. and Bosak, A. and Bouchet, J. and Bottin, F. and Dorado, B. and Paolasini, L. and Rennie, S. and Bell, C. and Springell, R. and Lander, G. H.},
    number = {3},
    month = {3},
    pages = {035004},
    volume = {5},
    publisher = {American Physical Society},
    url = {https://journals.aps.org/prmaterials/abstract/10.1103/PhysRevMaterials.5.035004},
    doi = {10.1103/PhysRevMaterials.5.035004},
    issn = {24759953},
    arxivId = {2009.03226}
}

@article{Zatterin2024AssessingSuperlattices,
    title = {{Assessing the Ubiquity of Bloch Domain Walls in Ferroelectric Lead Titanate Superlattices}},
    year = {2024},
    journal = {Physical Review X},
    author = {Zatterin, Edoardo and Ondrejkovic, Petr and Bastogne, Louis and Lichtensteiger, Céline and Tovaglieri, Ludovica and Chaney, Daniel A. and Sasani, Alireza and Sch{\"{u}}lli, Tobias and Bosak, Alexei and Leake, Steven and Zubko, Pavlo and Ghosez, Philippe and Hlinka, Jirka and Triscone, Jean Marc and Hadjimichael, Marios},
    number = {4},
    month = {11},
    pages = {041052},
    volume = {14},
    publisher = {American Physical Society},
    url = {https://journals.aps.org/prx/abstract/10.1103/PhysRevX.14.041052},
    doi = {10.1103/PhysRevX.14.041052},
    issn = {21603308}
}

@article{Bang2025High-energyFilms,
    title = {{High-energy diffuse X-ray scattering at ultra-small-angle grazing incidence for local structure study of single-crystalline thin films}},
    year = {2025},
    journal = {urn:issn:1600-5767},
    author = {Bang, Joohee and Strkalj, Nives and Sarott, Martin F. and Kholina, Yevheniia and Trassin, Morgan and Weber, Thomas},
    number = {4},
    month = {7},
    pages = {1417--1427},
    volume = {58},
    publisher = {International Union of Crystallography},
    url = {https://journals.iucr.org/paper?gue5004 https://journals.iucr.org/j/issues/2025/04/00/gue5004/},
    doi = {10.1107/S1600576725005837},
    issn = {16005767}
}

@article{Estandia2020Domain-Matching001,
    title = {{Domain-Matching Epitaxy of Ferroelectric Hf0.5Zr0.5O2 (111) on La2/3Sr1/3MnO3 (001)}},
    year = {2020},
    journal = {Crystal Growth {\&} Design},
    author = {Estand{\'{i}}a, Saul and Dix, Nico and Chisholm, Matthew F and Fina, Ignasi and S{\'{a}}nchez, Florencio},
    number = {6},
    pages = {3801--3806},
    volume = {20},
    publisher = {American Chemical Society (ACS)},
    doi = {10.1021/acs.cgd.0c00095},
    issn = {1528-7483}
}

@article{Song2020EpitaxialFilms,
    title = {{Epitaxial ferroelectric La-doped Hf0.5Zr0.5O2 thin films}},
    year = {2020},
    journal = {ACS Applied Electronic Materials},
    author = {Song, Tingfeng and Bachelet, Romain and Saint-Girons, Guillaume and Solanas, Raul and Fina, Ignasi and Sanchez, Florencio},
    number = {10},
    month = {10},
    pages = {3221--3232},
    volume = {2},
    publisher = {American Chemical Society},
    url = {/doi/pdf/10.1021/acsaelm.0c00560},
    doi = {10.1021/ACSAELM.0C00560/ASSET/IMAGES/LARGE/EL0C00560{\_}0012.JPEG},
    issn = {26376113}
}

@article{May2010QuantifyingFilms,
    title = {{Quantifying octahedral rotations in strained perovskite oxide films}},
    year = {2010},
    journal = {Physical Review B},
    author = {May, S. J. and Kim, J. W. and Rondinelli, J. M. and Karapetrova, E. and Spaldin, N. A. and Bhattacharya, A. and Ryan, P. J.},
    number = {1},
    month = {7},
    pages = {014110},
    volume = {82},
    publisher = {American Physical Society},
    url = {https://journals.aps.org/prb/abstract/10.1103/PhysRevB.82.014110},
    doi = {10.1103/PhysRevB.82.014110},
    issn = {10980121},
    arxivId = {1002.1317}
}

@article{Wadley2013ObtainingRadiation,
    title = {{Obtaining the structure factors for an epitaxial film using Cu X-ray radiation}},
    year = {2013},
    journal = {Journal of Applied Crystallography},
    author = {Wadley, P. and Crespi, A. and G{\'{a}}zquez, J. and Rold{\'{a}}n, M. A. and Garc{\'{i}}a, P. and Novak, V. and Campion, R. and Jungwirth, T. and Rinaldi, C. and Mart{\'{i}}, X. and Holy, V. and Frontera, C. and Rius, J.},
    number = {6},
    month = {12},
    pages = {1749--1754},
    volume = {46},
    publisher = {International Union of Crystallography},
    url = {https://journals.iucr.org/paper?rg5041 https://journals.iucr.org/j/issues/2013/06/00/rg5041/},
    doi = {10.1107/S002188981302414X},
    issn = {00218898}
}

@article{Nukala2020GuidelinesFilms,
    title = {{Guidelines for the stabilization of a polar rhombohedral phase in epitaxial Hf0.5Zr0.5O2 thin films}},
    year = {2020},
    journal = {Ferroelectrics},
    author = {Nukala, Pavan and Wei, Yingfen and De Haas, Vincent and Guo, Qikai and Antoja-Lleonart, Jordi and Noheda, Beatriz},
    number = {1},
    pages = {148--163},
    volume = {569},
    publisher = {Informa UK Limited},
    doi = {10.1080/00150193.2020.1791658},
    issn = {0015-0193}
}

@article{Ouyang2023StructuralHfO2,
    title = {{Structural stability and polarization analysis of rhombohedral phases of HfO2}},
    year = {2023},
    journal = {Applied Physics Letters},
    author = {Ouyang, Wenbin and Jia, Fanhao and Liu, Chang and Cheng, Xuli and Meng, Yaping and Gao, Ruiling and Picozzi, Silvia and Ren, Wei},
    number = {21},
    month = {11},
    pages = {212902},
    volume = {123},
    publisher = {American Institute of Physics Inc.},
    url = {/aip/apl/article/123/21/212902/2922847/Structural-stability-and-polarization-analysis-of},
    doi = {10.1063/5.0169911/2922847},
    issn = {00036951}
}

@article{ElBoutaybi2022StabilizationEnergy,
    title = {{Stabilization of the epitaxial rhombohedral ferroelectric phase in ZrO2 by surface energy}},
    year = {2022},
    journal = {Physical Review Materials},
    author = {El Boutaybi, Ali and Maroutian, Thomas and Largeau, Ludovic and Matzen, Sylvia and Lecoeur, Philippe},
    number = {7},
    month = {7},
    pages = {074406},
    volume = {6},
    publisher = {American Physical Society},
    url = {https://journals.aps.org/prmaterials/abstract/10.1103/PhysRevMaterials.6.074406},
    doi = {10.1103/PHYSREVMATERIALS.6.074406/FIGURES/3/MEDIUM},
    issn = {24759953},
    arxivId = {2111.05168}
}

@article{Wang2026GrainFerroelectrics,
    title = {{Grain boundary stabilization of fluorite ferroelectrics}},
    year = {2026},
    journal = {Nature Materials 2026},
    author = {Wang, Shiyu and Zhong, Hai and Song, Siyi and Gao, Ang and Zhang, Qinghua and Su, Dong and Jin, Kuijuan and Ge, Chen and Gu, Lin},
    month = {3},
    pages = {1--7},
    publisher = {Nature Publishing Group},
    url = {https://www.nature.com/articles/s41563-026-02533-6},
    doi = {10.1038/s41563-026-02533-6},
    issn = {1476-1122}
}

@book{Schroeder2019FerroelectricityDevices,
    title = {{Ferroelectricity in Doped Hafnium Oxide: Materials, Properties and Devices}},
    year = {2019},
    booktitle = {Electronic and Optical Materials},
    author = {Schroeder, Uwe and Hwang, Cheol Seong and Funakubo, Hiroshi},
    publisher = {Woodhead Publishing},
    url = {https://www.sciencedirect.com/science/article/pii/B9780081024300099903},
    isbn = {978-0-08-102430-0},
    doi = {https://doi.org/10.1016/B978-0-08-102430-0.09990-3}
}

@article{Estandia2021CriticalFilms,
    title = {{Critical effect of the bottom electrode on the ferroelectricity of epitaxial Hf0.5Zr0.5O2 thin films}},
    year = {2021},
    journal = {Journal of Materials Chemistry C},
    author = {Estand{\'{i}}a, Saúl and G{\`{a}}zquez, Jaume and Varela, María and Dix, Nico and Qian, Mengdi and Solanas, Raúl and Fina, Ignasi and S{\'{a}}nchez, Florencio},
    number = {10},
    month = {3},
    pages = {3486--3492},
    volume = {9},
    publisher = {The Royal Society of Chemistry},
    url = {https://pubs.rsc.org/en/content/articlehtml/2021/tc/d0tc05853j https://pubs.rsc.org/en/content/articlelanding/2021/tc/d0tc05853j},
    doi = {10.1039/D0TC05853J},
    issn = {2050-7534}
}

@article{Wolten1963DiffusionlessHafnia,
    title = {{Diffusionless Phase Transformations in Zirconia and Hafnia}},
    year = {1963},
    journal = {Journal of the American Ceramic Society},
    author = {Wolten, G M},
    number = {9},
    pages = {418--422},
    volume = {46},
    publisher = {Wiley},
    doi = {10.1111/j.1151-2916.1963.tb11768.x},
    issn = {0002-7820}
}

@article{Kaiser2023CrystalOxide,
    title = {{Crystal and Electronic Structure of Oxygen Vacancy Stabilized Rhombohedral Hafnium Oxide}},
    year = {2023},
    journal = {ACS Applied Electronic Materials},
    author = {Kaiser, Nico and Song, Young-Joon and Vogel, Tobias and Piros, Eszter and Kim, Taewook and Schreyer, Philipp and Petzold, Stefan and Valent{\'{i}}, Roser and Alff, Lambert},
    publisher = {American Chemical Society (ACS)},
    issn = {2637-6113}
}

@article{Zhang2024PhaseFatigue,
    title = {{Phase Transformation Driven by Oxygen Vacancy Redistribution as the Mechanism of Ferroelectric Hf0.5Zr0.5O2 Fatigue}},
    year = {2024},
    journal = {Advanced Electronic Materials},
    author = {Zhang, Zimeng and Craig, Isaac and Zhou, Tao and Holt, Martin and Flores, Raul and Sheridan, Evan and Inzani, Katherine and Huang, Xiaoxi and Nag, Joyeeta and Prasad, Bhagwati and Griffin, Sinéad M. and Ramesh, Ramamoorthy},
    number = {9},
    month = {9},
    pages = {2300877},
    volume = {10},
    publisher = {John Wiley {\&} Sons, Ltd},
    url = {/doi/pdf/10.1002/aelm.202300877 https://onlinelibrary.wiley.com/doi/abs/10.1002/aelm.202300877 https://advanced.onlinelibrary.wiley.com/doi/10.1002/aelm.202300877},
    doi = {10.1002/AELM.202300877},
    issn = {2199-160X}
}

@article{Katayama2016OrientationFilms,
    title = {{Orientation control and domain structure analysis of {\{}100{\}}-oriented epitaxial ferroelectric orthorhombic HfO2-based thin films}},
    year = {2016},
    journal = {Journal of Applied Physics},
    author = {Katayama, Kiliha and Shimizu, Takao and Sakata, Osami and Shiraishi, Takahisa and Nakamura, Shogo and Kiguchi, Takanori and Akama, Akihiro and Konno, Toyohiko J. and Uchida, Hiroshi and Funakubo, Hiroshi},
    number = {13},
    month = {4},
    volume = {119},
    publisher = {American Institute of Physics Inc.},
    issn = {10897550}
}

@article{Cho2024AtomicFilms,
    title = {{Atomic Layer Deposition of Epitaxial Ferroelectric Hf0.5Zr0.5O2 Thin Films}},
    year = {2024},
    journal = {Advanced Functional Materials},
    author = {Cho, Jung Woo and Song, Myeong Seop and Choi, In Hyeok and Go, Kyoung June and Han, Jaewoo and Lee, Tae Yoon and An, Chihwan and Choi, Hyung Jin and Sohn, Changhee and Park, Min Hyuk and Baek, Seung Hyub and Lee, Jong Seok and Choi, Si Young and Chae, Seung Chul},
    number = {24},
    month = {6},
    pages = {2314396},
    volume = {34},
    publisher = {John Wiley and Sons Inc},
    url = {/doi/pdf/10.1002/adfm.202314396 https://onlinelibrary.wiley.com/doi/abs/10.1002/adfm.202314396 https://advanced.onlinelibrary.wiley.com/doi/10.1002/adfm.202314396},
    doi = {10.1002/ADFM.202314396;JOURNAL:JOURNAL:10990712;ISSUE:ISSUE:DOI},
    issn = {16163028}
}

@article{Girard2019AESRF,
    title = {{A new diffractometer for diffuse scattering studies on the ID28 beamline at the ESRF}},
    year = {2019},
    journal = {urn:issn:1600-5775},
    author = {Girard, A. and Nguyen-Thanh, T. and Souliou, S. M. and Stekiel, M. and Morgenroth, W. and Paolasini, L. and Minelli, A. and Gambetti, D. and Winkler, B. and Bosak, A.},
    number = {1},
    month = {1},
    pages = {272--279},
    volume = {26},
    publisher = {International Union of Crystallography},
    url = {https://journals.iucr.org/paper?ay5525 https://journals.iucr.org/s/issues/2019/01/00/ay5525/},
    doi = {10.1107/S1600577518016132},
    issn = {1600-5775},
    pmid = {30655495}
}

@misc{2014CrysAlisPRO,
    title = {{CrysAlis PRO}},
    year = {2014},
    publisher = {Agilent Technologies Ltd.},
    address = {Yarnton, Oxfordshire, England}
}

@misc{2014CrystalMaker,
    title = {{CrystalMaker}},
    year = {2014},
    publisher = {CrystalMaker Software Ltd.},
    address = {Begbroke, Oxfordshire, England}
}

@article{Scott1988SwitchingMemories,
    title = {{Switching kinetics of lead zirconate titanate submicron thin‐film memories}},
    year = {1988},
    journal = {Journal of Applied Physics},
    author = {Scott, J F and Kammerdiner, ; L and Parris, ; M and Traynor, ; S and Ottenbacher, ; V and Shawabkeh, ; A and Oliver, ; W F and Scott, F and Kammerdiner, L and Parris, M and Traynor, S and Ottenbacher, V and Shawabkeh, A and Oliver, W F},
    number = {2},
    month = {7},
    pages = {787--792},
    volume = {64},
    publisher = {AIP Publishing},
    url = {/aip/jap/article/64/2/787/176417/Switching-kinetics-of-lead-zirconate-titanate},
    doi = {10.1063/1.341925},
    issn = {0021-8979}
}

\subsection*{Acknowledgments}
The authors thank Jacob Baas, Henk Bonder, Joost Zoestbergen and Mart Salverda for technical support. This publication is part of the TRICOLOR project, which is financed by the Dutch Research Council (NWO) through grant OCENW.M20.005 and by the Luxembourg National Research Fund (FNR) through grant INTER/NOW/20/15079143/TRICOLOR. This research was also partially supported by the Chinese Scholarship Council (CSC). We acknowledge the European Synchrotron Radiation Facility (ESRF) for provision of synchrotron radiation facilities under proposal number HC-5778. We also thank both Johannes Frey and Yves Watier as well other members of the ESRF sample environment group for their help developing and fabricating the custom furnace design employed. 

\subsection*{Conflict of Interest}
The authors declare there are no conflicts of interest.
\end{multicols}

\pagebreak

\subsection*{Supplementary Information}

\setcounter{figure}{0}
\renewcommand{\figurename}{Fig. S}
\vspace*{-3mm}
\begin{figure}[ht]
    \centering
        \includegraphics[width=1\textwidth]{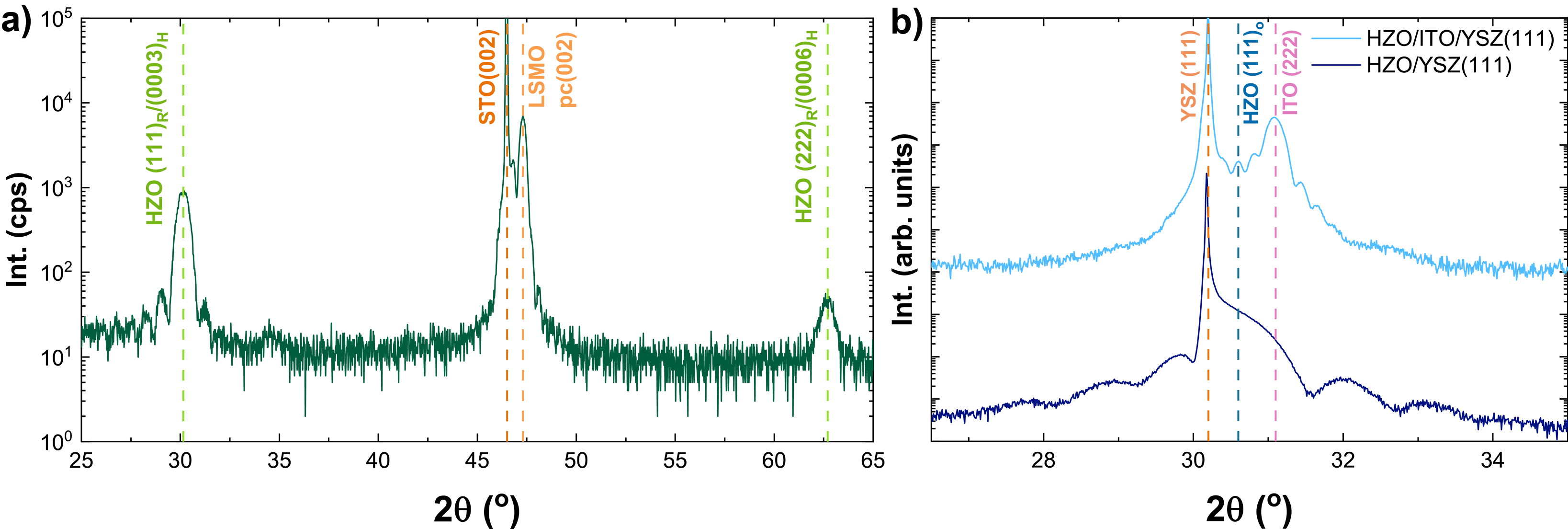}
        \hfill
    \caption{Specular \(\theta/2\theta\) scans for a) 10 nm HZO film grown on LSMO-buffered STO(100) and b) for a 10 nm HZO film grown on ITO-buffered (top) and directly on (bottom) YSZ(111) substrates.}
    \label{fig:SpecularScans}
\end{figure}

\begin{table}[ht]

    \caption{Lattice parameters obtained for the various hafnia/zirconia films of different thicknesses and stoichiometries grown on LSMO-buffered STO(001).}
    \begin{center}
    \begin{tabular}{|p{25mm}|p{115mm}|}
     \hline
    Sample & Lattice Parameters \\  
     \hline
         5 nm HfO$_2$ &
         \(a_{H}=b_{H}=6.15 Å,  c_{H}=8.98 Å, \alpha_{H}=\beta_{H}=90°, \gamma_{H}=120°\) \newline
         \( a_{R}=b_{R}=c_{R}=4.64 Å, \alpha_{R}=\beta_{R}=\gamma_{R}=82.93°\)
         \\ \hline
         10 nm YHO &  
         \(a_{H}=b_{H}=6.15 Å,  c_{H}=8.75 Å, \alpha_{H}=\beta_{H}=90°, \gamma_{H}=120°\) \newline
         \(a_{R}=b_{R}=c_{R}=4.60 Å, \alpha_{R}=\beta_{R}=\gamma_{R}=84.01^\circ\)
         \\ \hline
         5 nm HZO &  
        \(a_{H}=b_{H}=6.27 Å,  c_{H}=9.04 Å, \alpha_{H}=\beta_{H}=90°, \gamma_{H}=120°\) \newline
         \(a_{R}=b_{R}=c_{R}=4.71 Å, \alpha_{R}=\beta_{R}=\gamma_{R}=83.46°\)
         \\ \hline
         3 nm ZrO$_2$ &
         \(a_{\mathrm{H}}=b_{\mathrm{H}}=6.30 Å,  c_{\mathrm{H}}=8.99 Å, \alpha_{\mathrm{H}}=\beta_{\mathrm{H}}=90°, \gamma_{\mathrm{H}}=120^\circ \) \newline
         \(a_{\mathrm{R}}=b_{\mathrm{R}}=c_{\mathrm{R}}=4.71\ \text{\AA},\ \alpha_{\mathrm{R}}=\beta_{\mathrm{R}}=\gamma_{\mathrm{R}}=83.89^\circ\)
        \\    \hline
    \end{tabular}
    \end{center}
    \label{tab:LattParam}
\end{table}
\renewcommand{\thefigure}{S\arabic{figure}}
\begin{figure}[ht]
    \centering
        \includegraphics[width=1\textwidth]{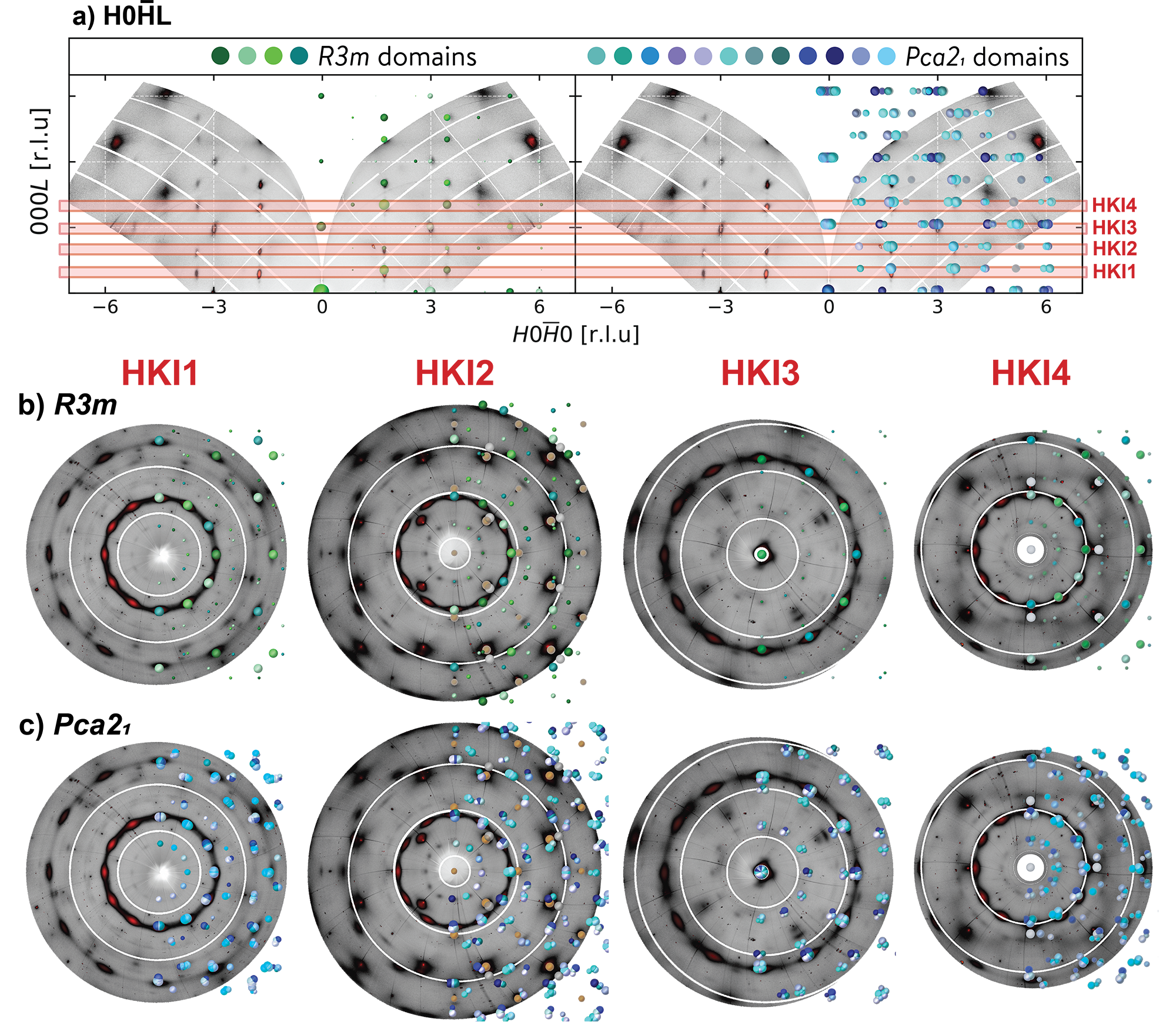}
        \hfill
    \caption{3D simulations of the \(R3m\) and \(Pca2_1\) phases for the a) in-plane (\(H0\bar HL\)) and b-c) various out-of-plane (\(HKIX\)) reconstructions of a 10 nm YHO film grown on LSMO-buffered STO(100).}
    \label{fig:HKXCuts}
\end{figure}

\begin{figure}[ht]
    \centering
        \includegraphics[width=1\textwidth]{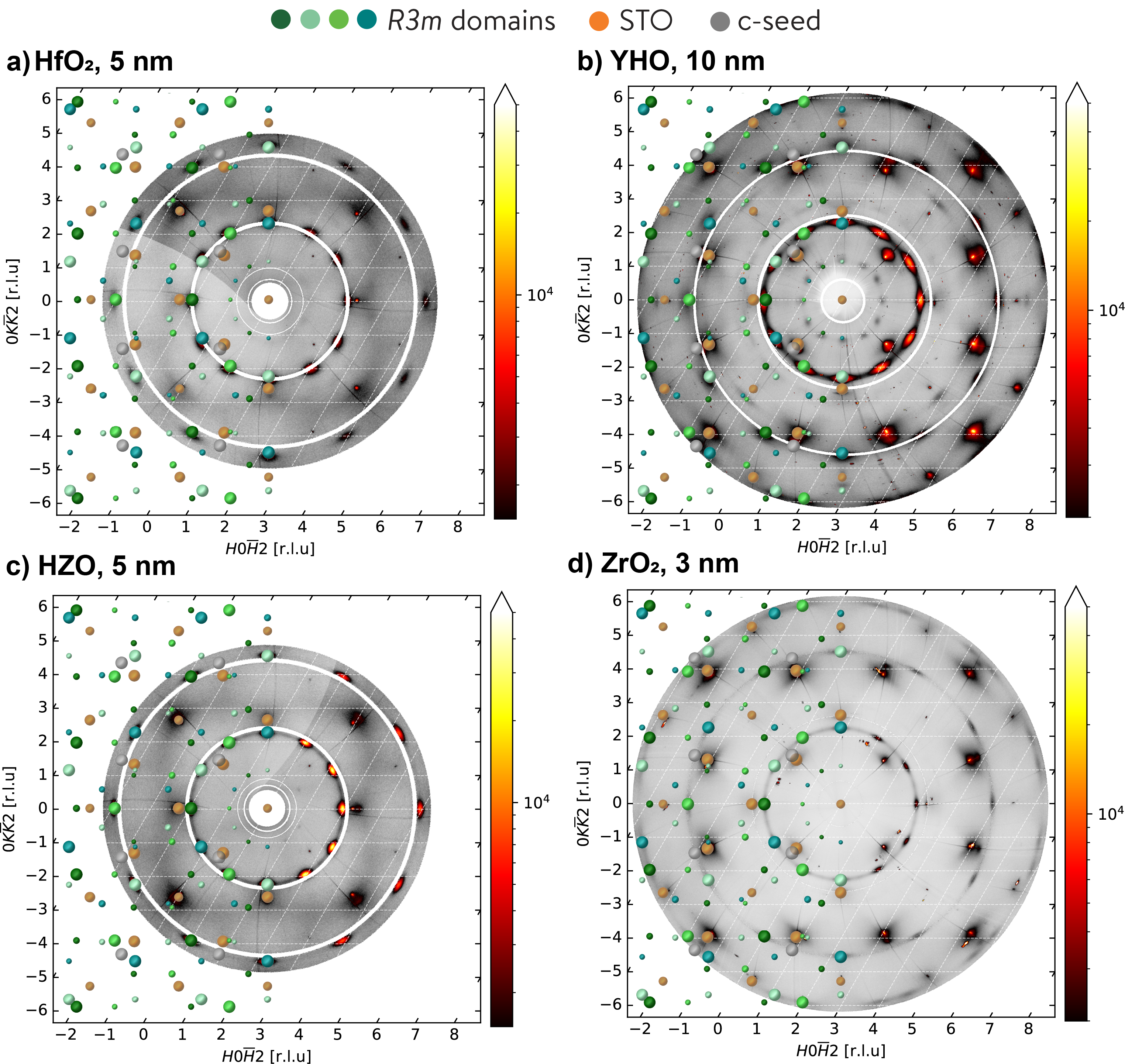}
        \hfill
    \caption{\(HKI2\) reconstructions for measurements on various films of different stoichiometries of hafnia/zirconia show good agreement with each other and can all be accurately simulated with the \(R3m\) structure. Note that for a) and c) each frame is integrated over an angular range of 0.50° for a time of $0.75\;\mathrm{s}$ and the volume of reciprocal space collected is smaller as only detector position 19° was used. For d) a thicker reciprocal space slice is integrated due to reduced film thickness.}
    \label{fig:HK2Stoich}
\end{figure}

\begin{figure}[ht]
    \centering
        \includegraphics[width=1\textwidth]{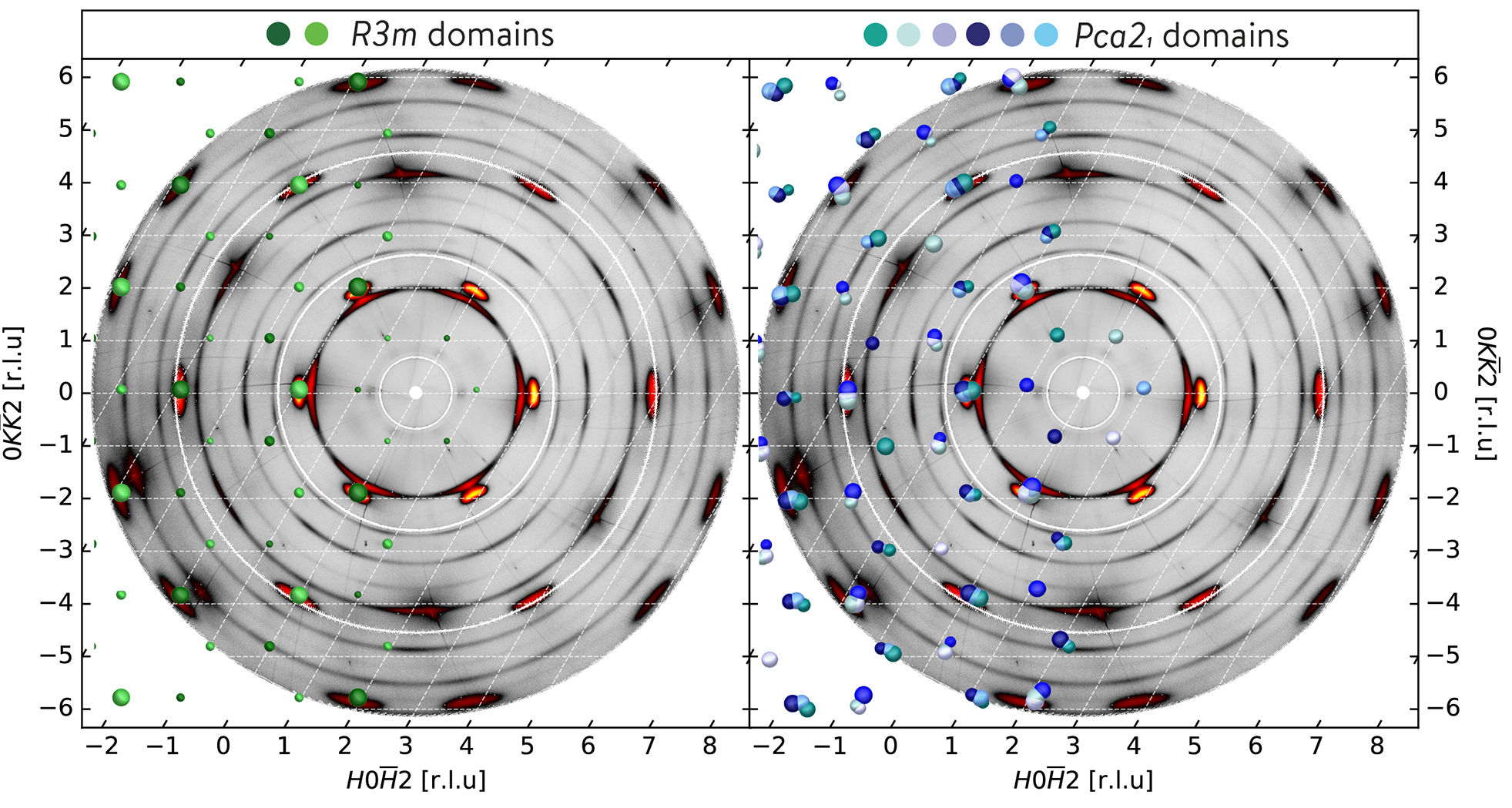}
        \hfill
    \caption{\(HKI2\) reconstruction for LSMO(27nm) \textbar \textbar HZO(4nm) \textbar \textbar LSMO(27nm) \textbar \textbar STO(110) with overlaid R-/OII-phase simulations. Note that the peaks and rings not accounted for in the simulations come from the LSMO layer grown above the HZO(111) layer corresponding to (110)-oriented epitaxial and polycrystalline growth, respectively.}
    \label{fig:STO110}
\end{figure}

\begin{figure}[ht]
    \centering
        \includegraphics[width=1\textwidth]{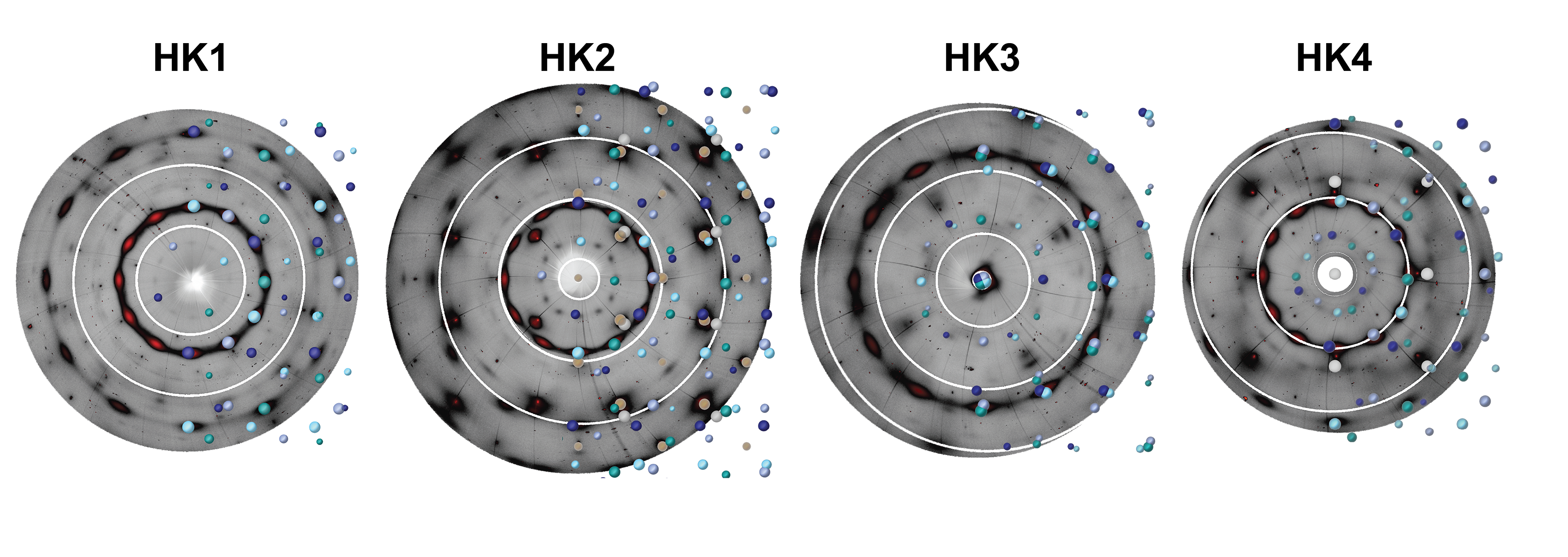}
        \hfill
    \caption{3D simulations of four 90° rotated domains of the \(Pca2_1\) phase for the various out-of-plane (\(HKIX\)) reconstructions of a 10 nm YHO film grown on LSMO-buffered STO(100).}
    \label{fig:4OIIIHKX}
\end{figure}

\begin{figure}[ht]
    \centering
        \includegraphics[width=1\textwidth]{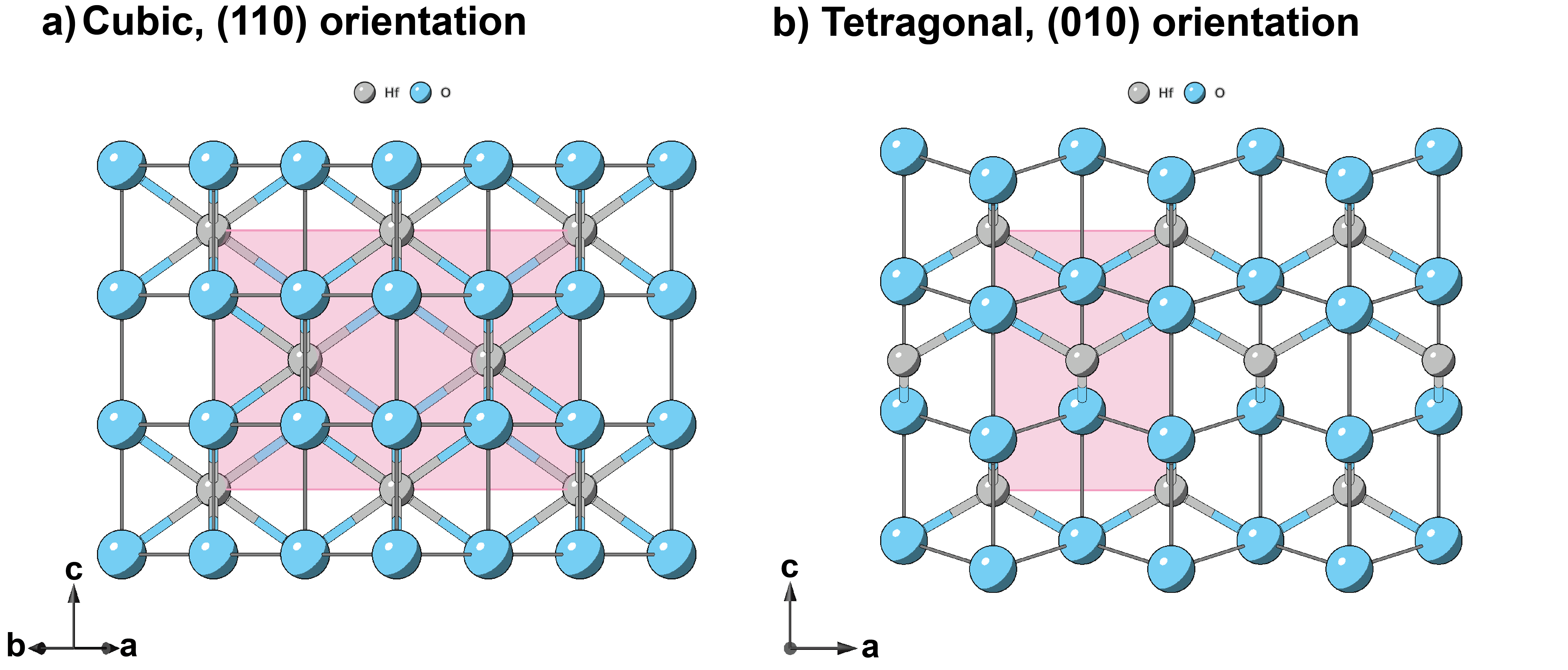}
        \hfill
    \caption{\emph{a)} (110)-orientation of cubic HfO$_2$ and \emph{b)} (010)-orientation of tetragonal HfO$_2$. The unit cell for each of the structures is indicated in pink. Since the heavier Hf atoms are in exactly in the same position, the only difference lies in the modulation of O atoms. Since the STEM data from Wei \emph{et al.} \cite{Wei2018AFilms} did not resolve the oxygen atoms, it is not possible to discern between the cubic and tetragonal structures through STEM alone.}
    \label{fig:SeedLayer}
\end{figure}

\begin{figure}[ht]
    \centering
        \includegraphics[width=0.9\textwidth]{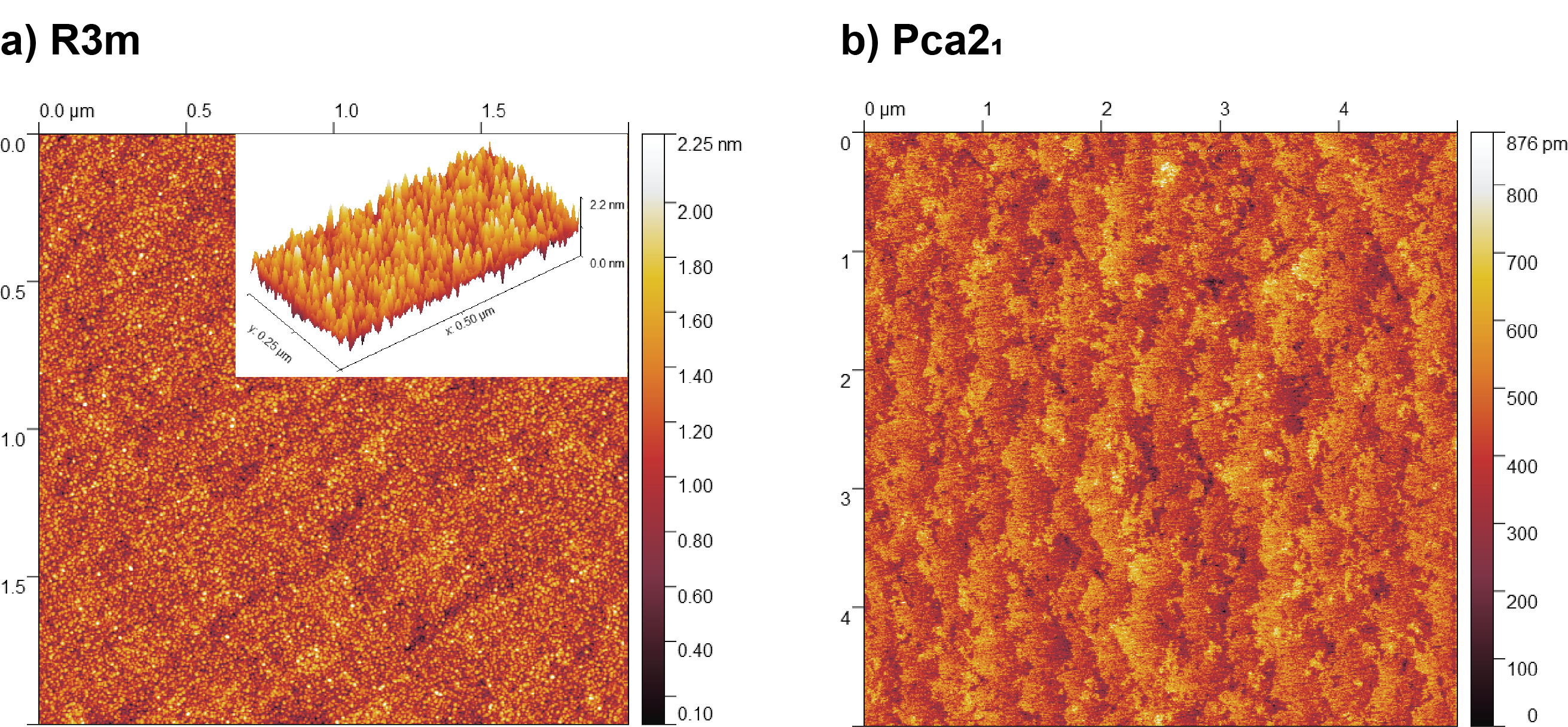}
        \hfill
    \caption{AFM topography measurements for \emph{a)} R-phase YHO and \emph{b)} OIII-phase YHO.}
    \label{fig:AFM}
\end{figure}

\begin{figure}[ht]
    \centering
        \includegraphics[width=1\textwidth]{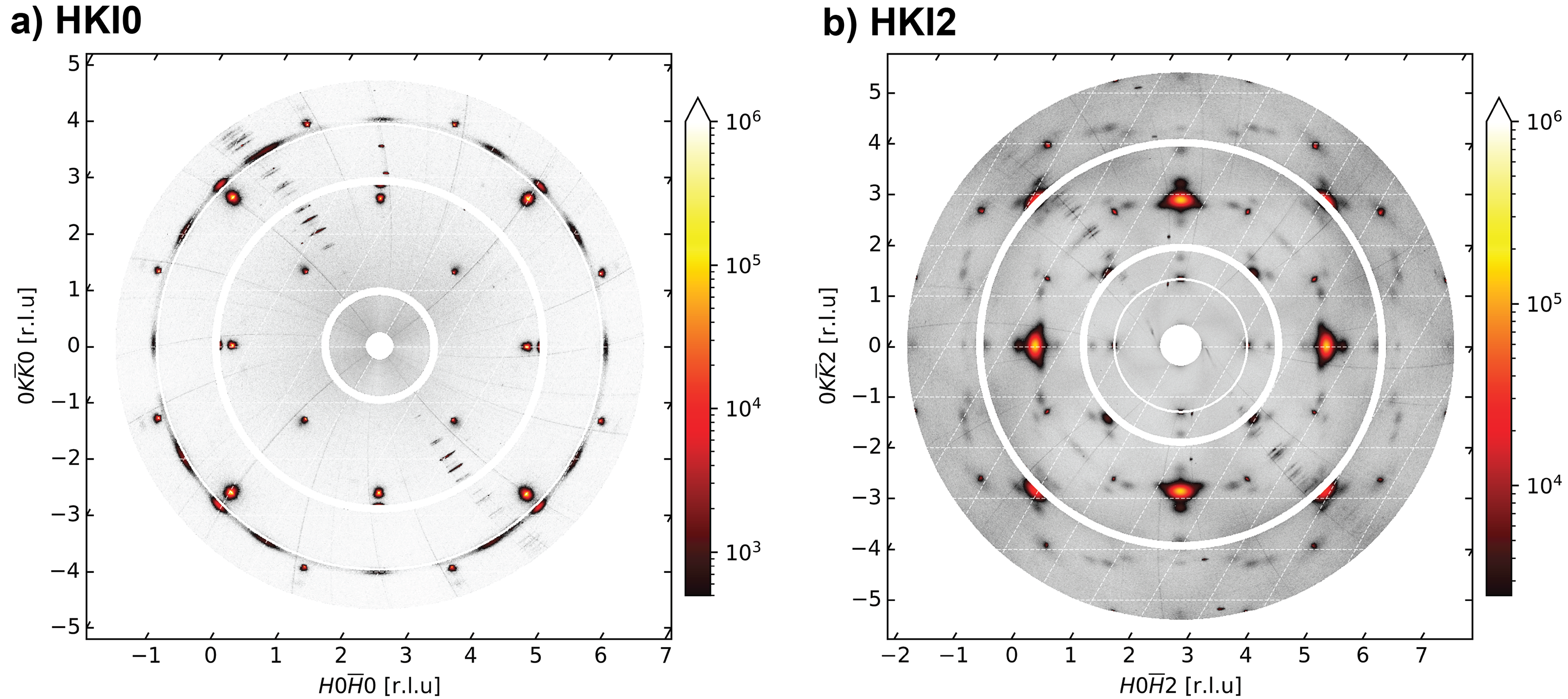}
        \hfill
    \caption{\emph{a)} HK0 and \emph{c)} HK2 reconstructions for a 10 nm YHO film grown directly on STO(001), which shows a mixture of cubic, monoclinic and rhombohedral phases.}
    \label{fig:DirectOnSTO}
\end{figure}

\begin{figure}[ht]
    \centering
        \includegraphics[width=1\textwidth]{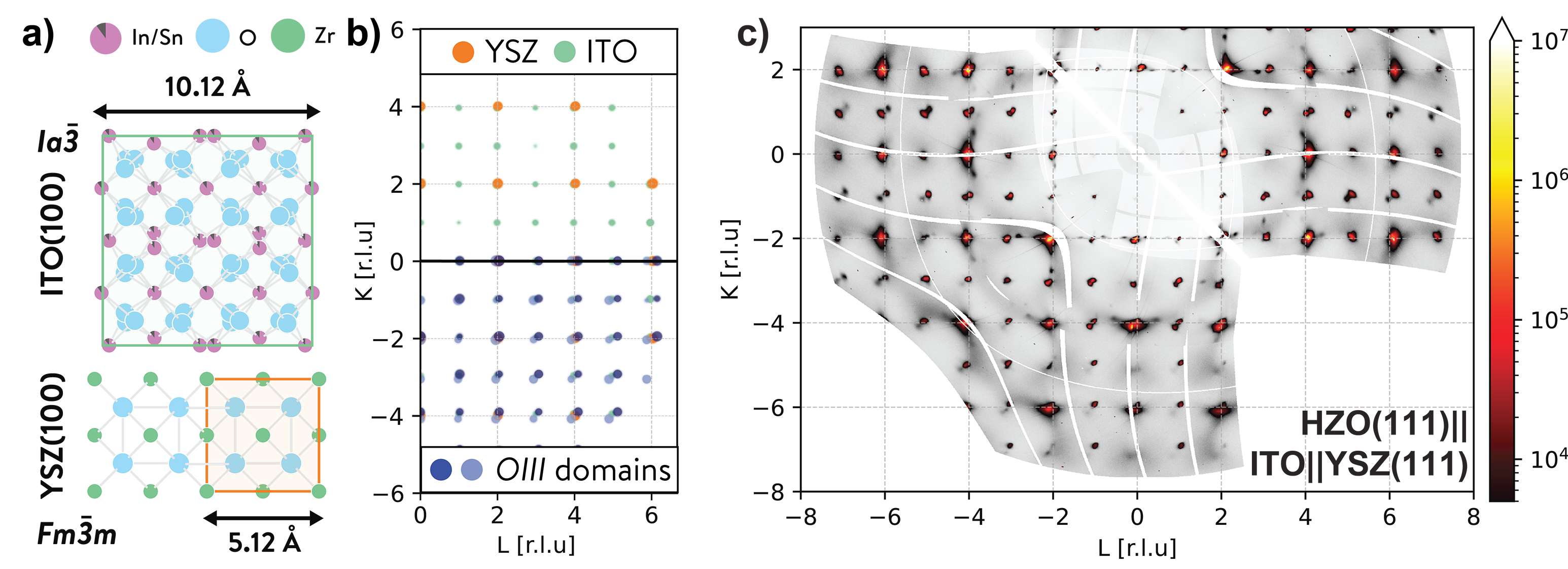}
        \hfill
    \caption{\emph{a)} Schematic diagram of the epitaxial match of ITO with YSZ. \emph{b)} Diffraction predictions for the 0KL cut of YSZ, ITO and OIII-HZO domains. c) Corresponding reconstruction for a 10 nm HZO film grown on ITO-buffered YSZ(111) substrate symmetrized by twofold rotation. Measurements conducted with incident wavelength of $0.73407\;\mathrm{\AA}$, integrating over angular range of $0.25\degree$ for $1\;\mathrm{s}$ maintaining a $12\;\mbox{keV}$ low energy threshold.}
    \label{fig:SCXRD_YSZ111}
\end{figure}

\begin{figure}[ht]
    \centering
        \includegraphics[width=1\textwidth]{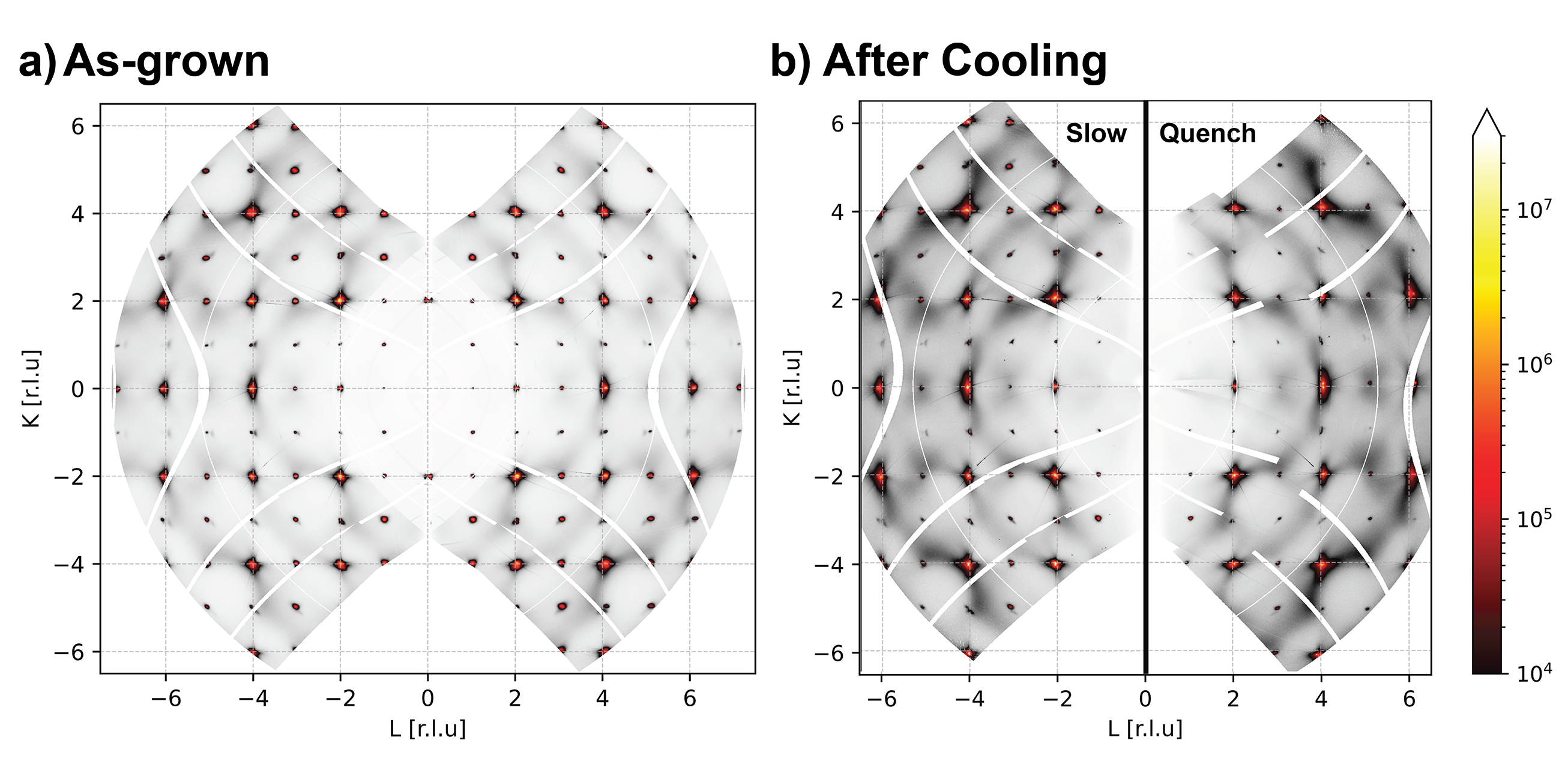}
        \hfill
    \caption{Comparison of the \(0KL\) reconstructions for a) as-grown YHO on YSZ(110) and b) the same film after cooling at 5°C/min (left) and quenching at approximately 300°C/min (right) from 1000°C.}
    \label{fig:Quench}
\end{figure}

\begin{figure}[ht]
    \centering
        \includegraphics[width=1\textwidth]{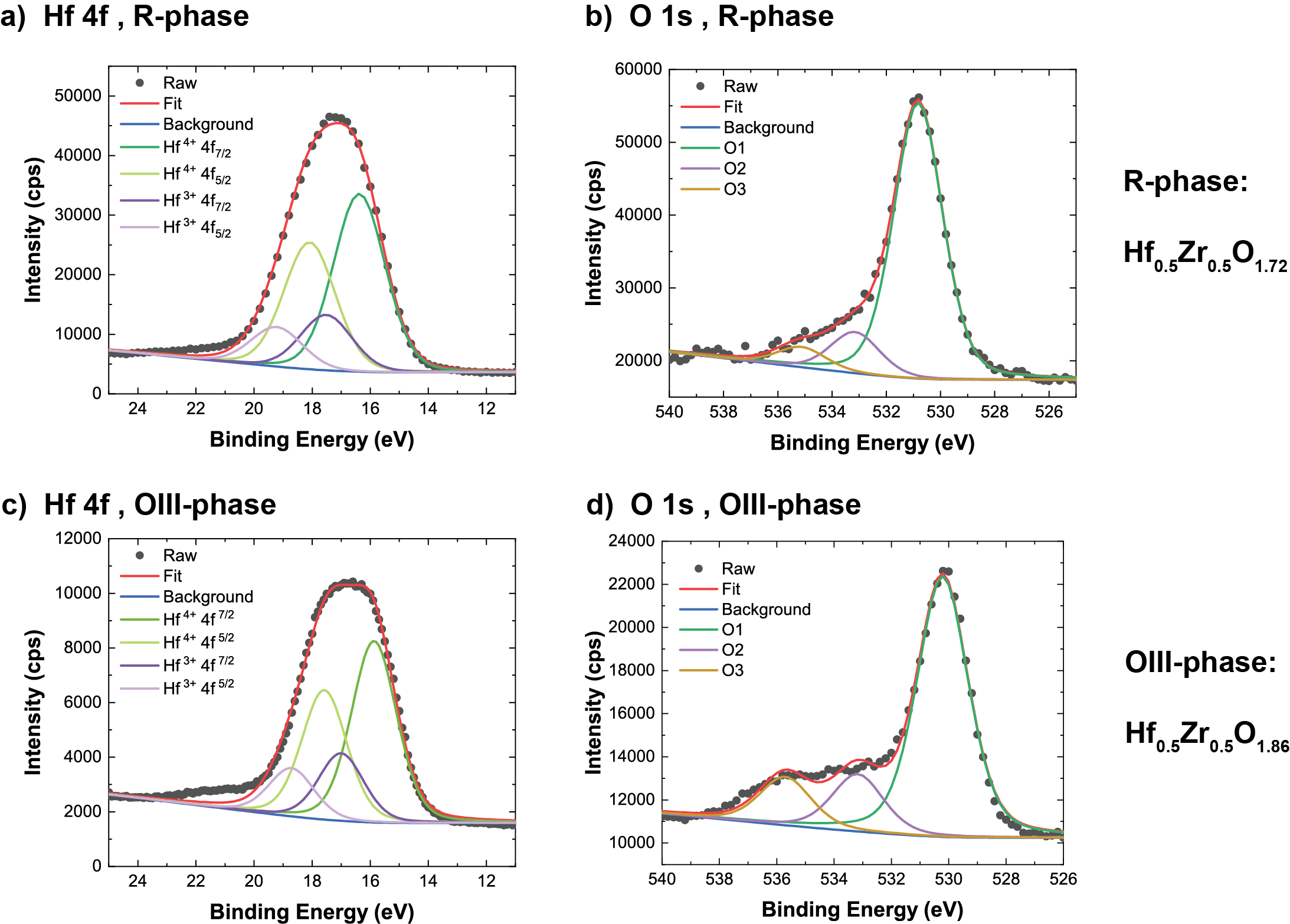}
        \hfill
    \caption{XPS measurements and the corresponding oxygen stoichiometry for \emph{a)} rhombohedral and \emph{b)} orthorhombic Hf$_{0.5}$Zr$_{0.5}$O$_n$ films grown under the same conditions. In both cases "O1" refers to the metal oxide oxygen peak (used for stoichiometry calculations) and "O2" and "O3" are non-lattice oxygen contributions (adsorbed species).}
    \label{fig:XPS}
\end{figure}

\pagebreak

\FloatBarrier

\begin{figure}[!tp]
    \centering
        \includegraphics[width=0.95\textwidth]{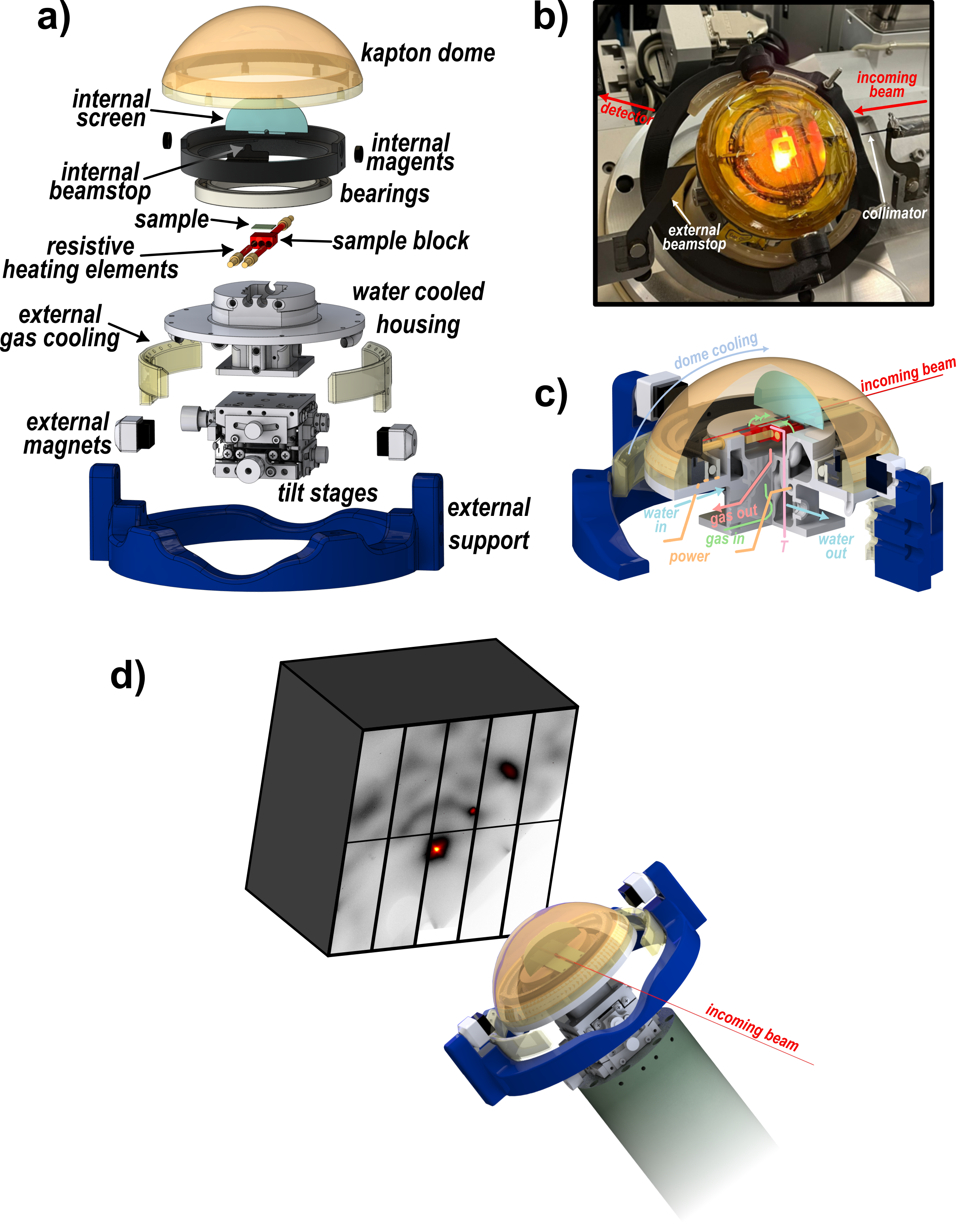}
    \caption{a) Exploded diagram of full furnace design. b) Photograph of furnace in operation. c) Cut out of main furnace body showing gas flow, water flow, thermocouple, power supply and dome cooling. d) Schematic representation of furnace setup in relation to detector}
    \label{fig:furnace_SM}
\end{figure}

\pagebreak

\FloatBarrier

\subsection*{Shadowless, ultra-low background, grazing incidence furnace}
  
To collect large volumes of reciprocal space containing low intensity Bragg reflections and/or diffuse x-ray scattering in grazing incidence geometry up to $1000\;\degree \mathrm{C}$ a custom furnace solution was developed and shown in Fig. \ref{fig:furnace_SM}. The solution is based on the following key components; i) a kapton dome to allow for any desired gas environment ii) a counter rotating beamstop/screen assembly to limit the resulting kapton scattering iii) three horizontal resistive heating elements supporting a ``free floating" heating stage separated from the main water cooled furnace body, iv) two additional tilt axis to align the film flat to the beam correcting for up to 3° misalignment in either axis.
 
Once aligned, the beamstop/screen assembly sits on a bearing and is held in place by a set of external magnets such that the entire assembly stays aligned with the beam during the measurement, removing both the front-side kapton scattering cone and stopping the straight through beam reaching the back-side of the dome thus preventing a second kapton scattering event. The main furnace body is water cooled in order to i) limit thermal expansion, ii) prevent the magnets rising above their curie temperature iii) negate the need for high temperature bearings iv) protect heating the tilt stages or goniometer mounting. The heating stage ``floats" in a cut-out supported by the heating elements themselves, which are hottest in the centre. The sample temperature is measured via a K-type thermocouple mounted in the heating block, 1 mm below the surface, and regulated remotely via a Eurotherm NanoDac regulator. To prevent softening and eventual failure of the kapton dome from radiative heating the dome is cooled by continuous gas flow delivered from two 3D-printed gas distribution segments mounted on the external support ring. Heating regulation is interlocked to both water and cooling gas flow such that the interruption of either cuts power to the heating elements preventing damage to the furnace. The sample environment gas is delivered via four symmetrically positioned outlets which flow directly over the sample, the positive pressure of the gas flow also keeps the dome inflated removing the need for a support structure and providing for ``shadowless" data collection. If one wishes to heat in an uncontrolled gas environment it is possible to run the system without the dome or internal beamstop/screen assembly allowing for higher sample throughput due to reduced alignment time but the increased collimator-external beamstop distance leads to higher background contribution from air scattering.

\end{document}